\documentclass[12pt]{article}
\pdfoutput=1
\usepackage{amsmath}
\usepackage{amssymb}
\usepackage{subfig}
\usepackage{graphicx}
\newcommand\blfootnote[1]{%
  \begingroup
  \renewcommand\thefootnote{}\footnote{#1}%
  \addtocounter{footnote}{-1}%
  \endgroup
}
\numberwithin{equation}{section}
\begin{document}

\vspace*{-1.5cm}
\thispagestyle{empty}
\begin{flushright}
AEI-2012-033
\end{flushright}
\vspace*{2.5cm}
\begin{center}
{\Large
{\bf The limit of $\boldsymbol{N=(2,2)}$ superconformal minimal models}}
\vspace{2.5cm}

{\large Stefan Fredenhagen, Cosimo Restuccia, Rui Sun}
\blfootnote{{\tt E-mail: FirstName.LastName@aei.mpg.de}}

\vspace*{0.5cm}

Max-Planck-Institut f{\"u}r Gravitationsphysik\\
Albert-Einstein-Institut\\
Am M{\"u}hlenberg 1\\
14476 Golm, Germany\\
\vspace*{3cm}

{\bf Abstract}
\end{center}

The limit of families of two-dimensional conformal field theories has
recently attracted attention in the context of AdS/CFT dualities. In
our work we analyse the limit of $N= (2,2)$ superconformal minimal models when
the central charge approaches $c=3$. The limiting theory is a
non-rational $N= (2,2)$ superconformal theory, in which there is a
continuum of chiral primary fields. We determine the spectrum of the
theory, the three-point functions on the sphere, and the disc one-point
functions.

\newpage

\tableofcontents

\section{Introduction}

The analysis and construction of non-rational conformal field theories (CFTs)
in two dimensions is a highly non-trivial task. On the other hand,
rational CFTs are well investigated and understood. Some non-rational
theories can be constructed as limits of rational theories, and their
properties can be inferred from our knowledge about rational
models. The first example of such a limiting theory is the
Runkel-Watts theory~\cite{Runkel:2001ng} at central charge $c=1$ that arises as the 
limit of Virasoro minimal models. Similar constructions have been
considered for $W_{n}$ minimal models~\cite{Fredenhagen:2010zh}, and supersymmetric $N=1$
minimal models~\cite{Fredenhagen:2007tk}. A different approach towards
taking limits of conformal field theories is discussed in~\cite{Roggenkamp:2003qp}.

It is an obvious question whether such a construction is also possible
for the $N= (2,2)$ supersymmetric minimal models. For several reasons
this is far from being a straightforward generalisation of the known
cases. Firstly, all other examples are constructed as diagonal coset
models of the form $\frac{\mathfrak{g}_{k}\oplus
\mathfrak{g}_{\ell}}{\mathfrak{g}_{k+\ell}}$ where the level $k$ is
sent to infinity. On the other hand, the $N= (2,2)$ Grassmannian
Kazama-Suzuki models~\cite{Kazama:1989qp}, of which the $N= (2,2)$
minimal models are the simplest example, have a coset description as
$su(n)/u(n-1)$, so their structure is different. Secondly, for the
diagonal cosets it is
known~\cite{Zamolodchikov:1987ti,Ludwig:1987gs,Lukyanov:1990tf} that
there are renormalisation group (RG) flows that connect theories with
different levels $k$ (triggered by the $(1,1;\text{Adjoint})$ field).
These flows become short\footnote{with respect to the Zamolodchikov
metric~\cite{Zamolodchikov:1986gt}} for large levels~$k$ (being
accessible to conformal perturbation theory), and the models come
closer in the space of theories, therefore one would intuitively expect a
``convergence'' to a limiting theory. For the $N= (2,2)$ minimal
models there are also RG flows connecting different models, but they
are not accessible to conformal perturbation
theory~\cite{Cvetic:1989qv,LeafHerrmann:1990db} and are not short with
respect to the Zamolodchikov metric, so that the theories do not seem
to approach a limiting point in theory space. From this point of view,
one might even doubt that a limit of $N= (2,2)$ minimal models for
large levels can be defined.

In this article we are investigating precisely this question. We
analyse the spectrum of the $N= (2,2)$ minimal models in the limit of
large levels $k$ for which the central charge approaches $c=3$. In the
Neveu-Schwarz sector we find primary fields $\Phi_{q,n}$ with a
continuous charge $0<|q|<1$ and a discrete label $n=0,1,\dotsc$. The
fields with label $n=0$ are chiral or anti-chiral primaries. By taking
the limit of the known three-point functions of minimal models, we
show that the fields in the limit theory have well-defined and
non-trivial three-point functions. We also can define two classes of
boundary conditions, a discrete one labelled by an integer $M$ and a
continuous one labelled -- similarly to the fields -- by a continuous
parameter $Q$ and a discrete parameter $N$, and we determine the disc
one-point functions. One might still wonder whether the limit theory
is fully consistent, but the results so far indicate that it is well
behaved. It would be interesting to check that the resulting theory
satisfies crossing symmetry.

Limits of conformal field theories also appear in the context of
AdS/CFT dualities for higher spin gravity theories. Starting from the
observation that the asymptotic symmetry of a higher spin gravity
theory on AdS$_{3}$ is given by a
W-algebra~\cite{Henneaux:2010xg,Campoleoni:2010zq}, Gaberdiel and
Gopakumar proposed a certain limit of $W_{n}$ minimal models as the
corresponding CFT dual~\cite{Gaberdiel:2010pz}. In this limit, both
the level $k$ and the label $n$ are sent to infinity such that the 't
Hooft coupling $\lambda =\frac{n}{k+n}$ is kept fixed. This proposal
was generalised to $N= (2,2)$ superconformal theories
in~\cite{Creutzig:2011fe,Candu:2012jq}. In this context, our approach
to send $k$ to infinity in a given theory (with fixed $n$) is related
to the situation where the 't Hooft coupling is zero.
\smallskip

The paper is organised as follows. In section~2 we consider the
behaviour of the spectrum of minimal models in the limit. We define
fields in the limit theory and show that they have sensible two-point
functions.  In section~3 we compute the limit of the three-point
function, the necessary technical and computational details are
collected in three appendices. Section~4 discusses boundary conditions
and disc one-point functions. In section~5 we deal with the question
whether we can define further fields of charge zero in the limit
theory, and we conclude in section~6.

\section{The spectrum}
\label{sec:spectrum}
In this section we will analyse the spectrum of the limit theory. We
start by reviewing some facts about minimal models, and then study
their spectrum for large levels and define the corresponding fields 
in the limit theory.

\subsection{Minimal models} 

The $N= (2,2)$ superconformal minimal models\footnote{For an
introduction see e.g.\ the
textbooks~\cite{PolchinskiBookII:1998,Blumenhagen:book}.} come in a
family parameterised by a positive integer $k$ with central charges
\begin{equation}
c=3\frac{k}{k+2}\ .
\end{equation}
They possess a discrete spectrum. The unitary representations of the
bosonic subalgebra of the $N=2$ superconformal algebra are labelled by
three integers $(l,m,s)$, where
\begin{equation}
0\leq l \leq k\quad ,\quad m\equiv m+2k+4 \quad ,\quad s\equiv s+4 \ .
\end{equation}
Only those triples $(l,m,s)$ are allowed for which $l+m+s$ is even,
and triples are identified according to the relation
\begin{equation}\label{cosetident}
(l,m,s) \equiv (k-l,m+k+2,s+2) \ .
\end{equation}
The conformal weight and the $U (1)$ charge of the vectors in a
representation $\mathcal{H}_{(l,m,s)}$ are given by
\begin{align}
h & \in h_{l,m,s} + \mathbb{N}  &  h_{l,m,s} &= \frac{l
(l+2)-m^{2}}{4 (k+2)}+\frac{s^{2}}{8}\\
q & \in q_{m,s} + 2\mathbb{Z} & q_{m,s} & =
-\frac{m}{k+2}+\frac{s}{2} \ .
\end{align}
We consider models with a diagonal spectrum, i.e.\ with equal left-
and right-moving weights, $\bar{h}=h$, and charges, $\bar{q}=q$, of
the ground states. The conformal weight and the $U (1)$ charge of the
ground states of $\mathcal{H}_{(l,m,s)}$ are exactly given by
$h_{l,m,s}$ and $q_{m,s}$ (without integer shifts) if the labels
satisfy
\begin{equation}
|m-s| \leq l \ ,
\end{equation}
which is sometimes called the standard range. Contrary to some claims
in the literature, the identification rule~\eqref{cosetident} does not
allow one in general to map a given triple into the standard range.
Exceptions are provided by superdescendants of chiral primary or
Ramond ground states (e.g.\ $(0,0,2)\equiv (k,k+2,0)$ cannot be mapped
to the standard range).

Representations with even $s$ belong to the Neveu-Schwarz sector. The
direct sum $\mathcal{H}_{(l,m,0)}\oplus \mathcal{H}_{(l,m,2)}$
constitutes a representation of the full superconformal algebra. The
primary fields  $\phi_{l,m}$ with respect to the superconformal
algebra are then labelled by a pair of integers $(l,m)$, where
\begin{equation}\label{conditions_on_l_and_m}
0\leq l \leq k \quad ,\quad |m|\leq k \quad , \quad l+m\ \text{even} \ .
\end{equation}
Their conformal weights and $U (1)$-charges are given by
\begin{align}
h_{l,m}=h_{l,m,0} & = \frac{l (l+2)-m^{2}}{4 (k+2)} \label{def_hlm}\\
q_{m,0} &= -\frac{m}{k+2} \ .
\end{align}
The chiral primary fields are those with $m=-l$ obeying
$h_{l,-l}=q_{l,-l}/2$, the anti-chiral primary states have $m=l$.

Representations with odd $s$ belong to the Ramond sector. The Ramond
ground states have labels $(l,l+1,1)$ with weight and charge given by
\begin{align}
h_{l,l+1,1} & = \frac{1}{8}-\frac{1}{4 (k+2)}\\
q_{l+1,1} & = \frac{1}{2} - \frac{l+1}{k+2} \ ,
\end{align}
the corresponding field will be denoted by $\psi_{l}^{0}$. The full
representation of the superconformal algebra built on such Ramond
ground states is then $\mathcal{H}_{(l,l+1,1)}\oplus
\mathcal{H}_{(l,l+1,-1)}$. The other Ramond representations of the
superconformal algebra are given by the sum
$\mathcal{H}_{(l,m,1)}\oplus \mathcal{H}_{(l,m,-1)}$ with $|m|\leq
l-1$. The ground states in the two summands have the same conformal
weight and differ by $1$ in the $U (1)$ charge,
\begin{align}
h_{l,m,\pm 1} & = \frac{l (l+2)-m^{2}}{4 (k+2)} + \frac{1}{8} \\
q_{m,\pm 1} & = -\frac{m}{k+2} \pm \frac{1}{2} \ .  
\end{align}
We denote the corresponding two fields by $\psi_{l,m}^{\pm}$. 

\subsection{Taking the limit}

We want to take the limit $k\to \infty$, and analyse what happens to
the spectrum. Let us first consider the primary states in the
Neveu-Schwarz sector. When the level $k$ becomes large, the spectrum
of $U (1)$-charges becomes continuous in the range $-1<q<1$. We want
to keep the $U (1)$ charge and the conformal weight fixed in the
limit. For a fixed charge $q$ we have to scale $m$ with $k$ such that
\begin{equation}
m \approx -q (k+2)\ .
\end{equation}
On the other hand, the label $l$ is determined by $h_{l,m,0}$ and
$q_{m,0}$ by 
\begin{equation}\label{l_intermsof_hq}
l = \sqrt{(k+2)^{2}q_{m,0}^{2} + 4 (k+2)h_{l,m,0}+1}-1 \ .
\end{equation}
Keeping $q_{m,0} \approx q \not= 0$ and $h_{l,m,0}\approx h$ fixed, the
label $l$ scales as 
\begin{equation}\label{l_in_the_limit}
l = |m| + 2 \frac{h}{|q|} -1 + \mathcal{O} (1/k) \ .
\end{equation}
The label $l$ thus differs from the linearly growing $|m|$ only by a
fixed finite number, which has to be an even integer
(see~\eqref{conditions_on_l_and_m}),
\begin{equation}\label{relation_l_and_m}
l = |m| + 2n \quad ,\quad n=0,1,2,\dotsc \ .
\end{equation}
Whereas $|q|$ can take any value between $0$ and $1$, we see by
comparing~\eqref{l_in_the_limit} and~\eqref{relation_l_and_m} that the
ratio $h/|q|$ can only take discrete values,
\begin{equation}
h_{n} (q) = (2n+1) |q|/2 \ ,
\end{equation}
and $n=0$ corresponds to chiral primary and anti-chiral primary fields.

\begin{figure}
\begin{center}
\includegraphics[width=7.5cm]{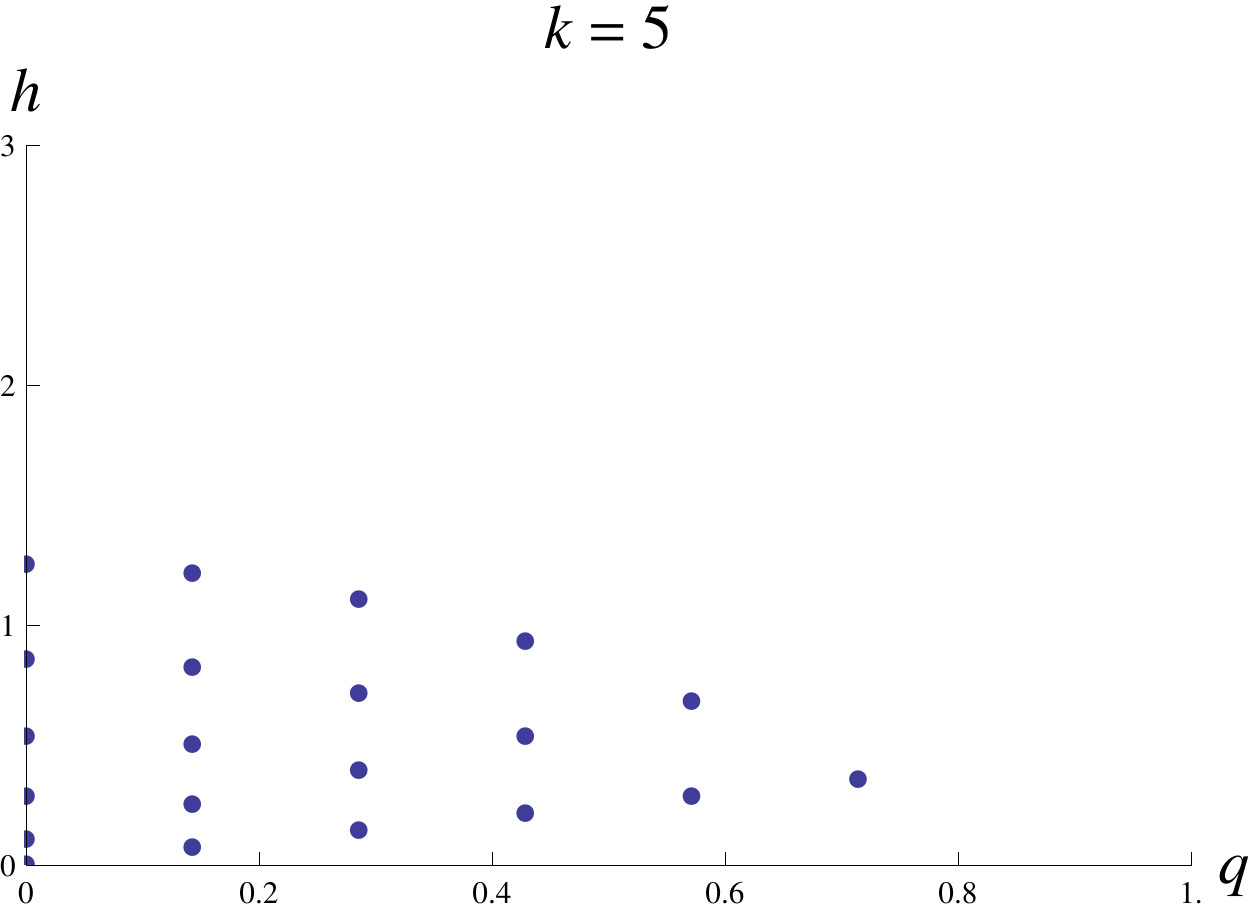}\ \  
\includegraphics[width=7.5cm]{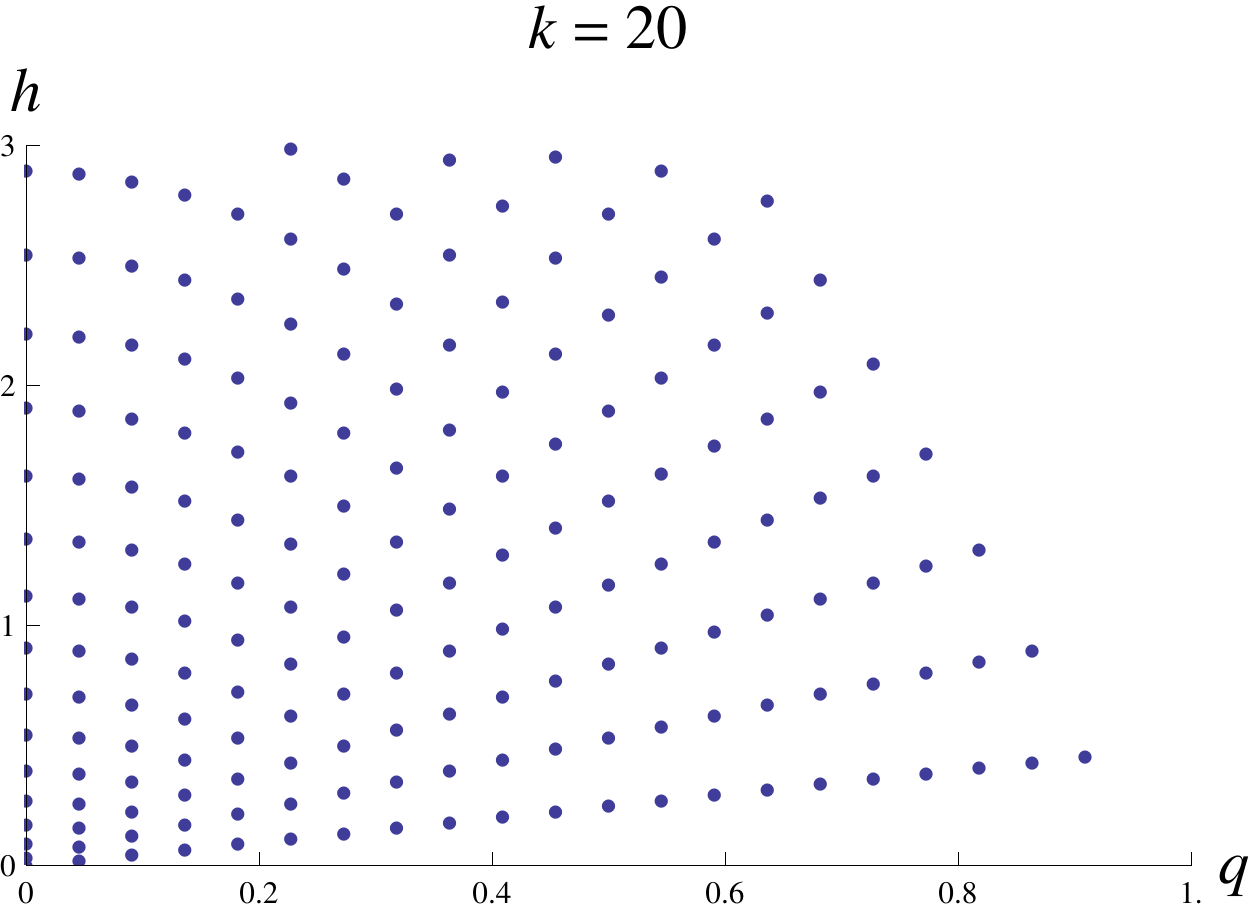}\\[8mm]
\includegraphics[width=7.5cm]{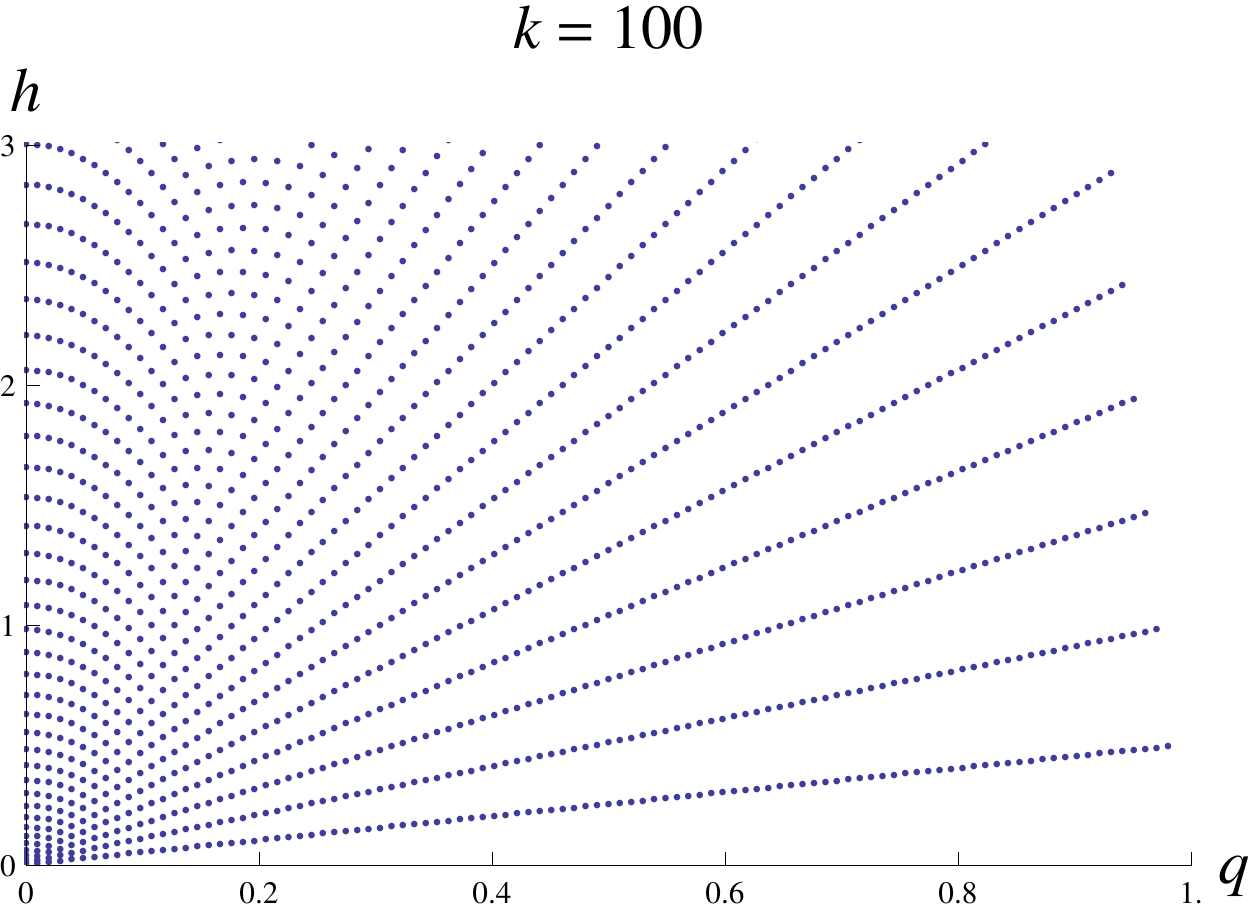}\ \ 
\includegraphics[width=7.5cm]{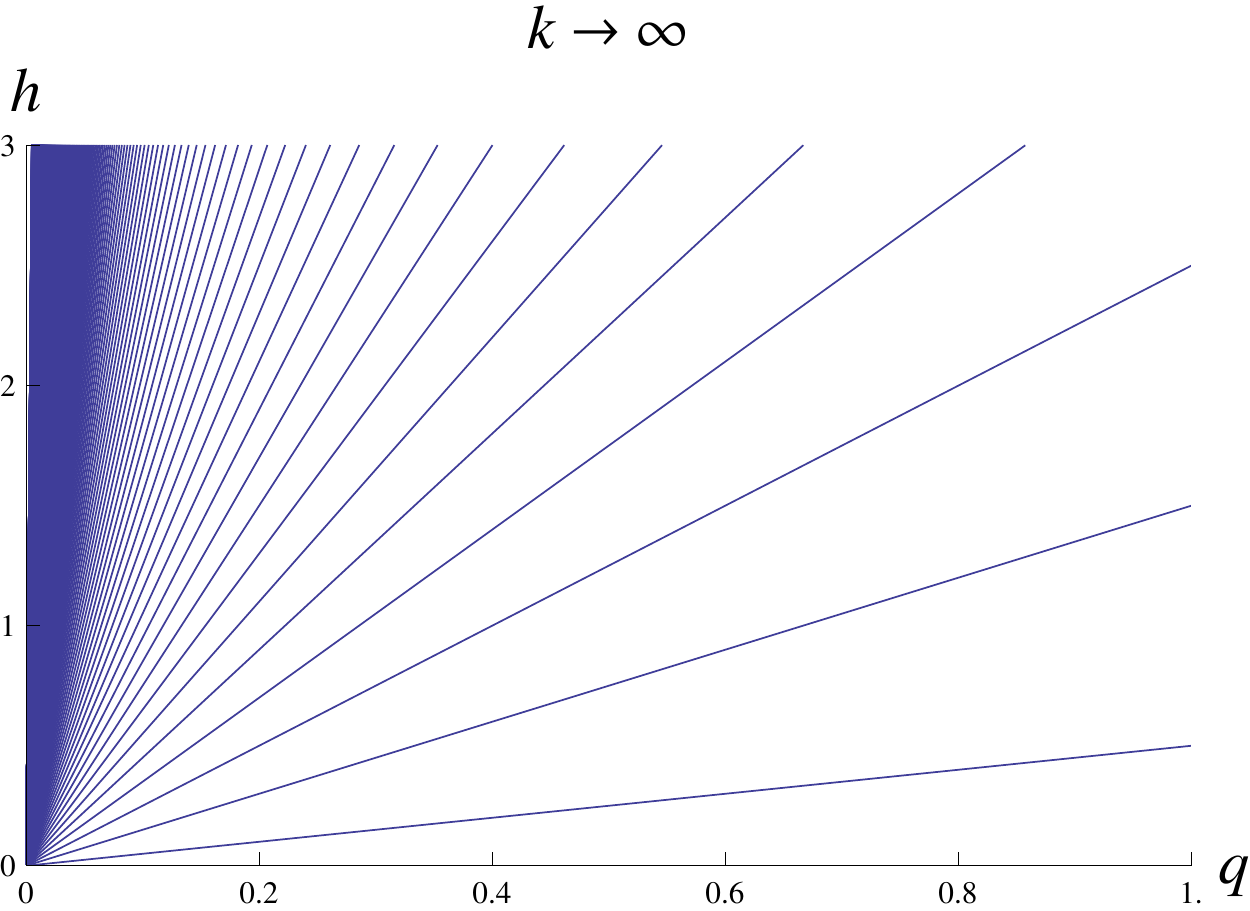}
\end{center}
\caption{\label{fig:spectrum}Behaviour of the spectrum of primary
fields in the Neveu-Schwarz sector for large
levels $k$: when one plots the values of the conformal weight $h$
and of the $U (1)$ charge $q$ as dots in the $h$-$q$-plane, one observes
that the points assemble along straight lines starting from the
origin. Notice that we only plotted the points corresponding to
positive charge $q$ (the negative charged part is just the mirror
picture) and we truncated the conformal weights by $h\leq 3$.}
\end{figure}

In the $h$-$q$-plane, the Neveu-Schwarz spectrum is thus concentrated
on lines going through the origin (see figure~\ref{fig:spectrum}), and
the fields $\Phi_{q,n}$ are labelled by their continuous $U
(1)$-charge $q$ and a discrete label $n$.
\smallskip

By a similar analysis we find in the Ramond sector on the one hand the
Ramond ground states leading to fields $\Psi^{0}_{q}$ with
$h=\frac{1}{8}$ and $-\frac{1}{2}<q<\frac{1}{2}$ built from fields
$\psi^{0}_{l}$ with $l\approx (k+2)(\frac{1}{2}-q)$. In addition
there are the fields $\Psi^{\pm}_{q,n}$ with $-\frac{1}{2}<\pm q
<\frac{3}{2}$ and 
\begin{equation}
h^{\pm}_{n} (q) =\frac{1}{8} + n \left| q\mp \frac{1}{2}\right| \ .
\end{equation}
They are obtained from fields $\psi^{\pm}_{l,m}$ with $l=|m|+2n-1$
and $m\approx - (k+2) \left( q\mp \frac{1}{2}\right)$.

\subsection{Fields and correlators}

We now want to become more precise about how the limit of the fields is
taken. We focus here on the Neveu-Schwarz sector, the construction in
the Ramond sector is analogous. 

For the fields $\Phi_{q,n}$ with $0<|q|<1$ we proceed as follows. We
first define averaged fields,
\begin{equation}
\Phi_{q,n}^{\epsilon ,k} = \frac{1}{|N (q,\epsilon ,k)|} 
\sum_{\substack{m \in N (q,\epsilon ,k)\\ l = |m|+2n}}
\phi_{l,m} \ ,
\end{equation} 
where the set $N (q,\epsilon ,k)$ contains all labels $m$ such that the
corresponding charge $q_{m}$ is close to $q$, more precisely
\begin{equation}
N (q,\epsilon ,k) = \left\{m \left| q-\frac{\epsilon}{2}< -\frac{m}{k+2} <
q+\frac{\epsilon}{2}\right. \right\} \ .
\end{equation}
The cardinality of the set is
\begin{equation}
|N (q,\epsilon ,k)| = \epsilon (k+2) + \mathcal{O} (1) \ .
\end{equation}
We assume that $\epsilon$ is small enough such that $|q|\pm
\frac{\epsilon}{2}$ is still between $0$ and $1$.

The correlator of fields in the limit theory is then defined as
\begin{equation}\label{defofcorrelators}
\langle \Phi_{q_{1},n_{1}} (z_{1},\bar{z}_{1}) \dotsb \Phi_{q_{r},n_{r}} (z_{r},\bar{z}_{r})\rangle
= \lim_{\epsilon \to 0} \lim_{k\to \infty} 
\beta (k)^{2} \alpha (k)^{r}\langle \Phi_{q_{1},n_{1}}^{\epsilon ,k}
(z_{1},\bar{z}_{1})\dotsb \Phi_{q_{r},n_{r}}^{\epsilon ,k} (z_{r},\bar{z}_{r})\rangle \ ,
\end{equation}
where $\beta (k)^{2}$ is a factor that can be used to change the
normalisation of the correlator in the limit (which corresponds to a
rescaling of the vacuum by a factor $\beta (k)$), while $\alpha (k)$
is a factor that is used to change the normalisation of the fields
while taking the limit.\footnote{We could allow $\alpha$ to depend
also on the field labels $q,n$, but it will turn out that this is not
necessary.} The $k$-dependence of $\alpha$ and $\beta$ are determined
such that we obtain finite correlators in the limit. Obviously we need
at least two correlators with a different number of fields to
determine the $k$-dependence of both factors $\alpha$ and $\beta$.

Let us now analyse the two-point function. We normalise the fields in the
minimal models such that
\begin{equation}
\langle \phi_{l_{1},m_{1}} (z_{1}) \phi_{l_{2},m_{2}} (z_{2})\rangle
= \delta_{l_{1},l_{2}}\delta_{m_{1},-m_{2}}
\frac{1}{|z_{1}-z_{2}|^{4h_{l_{1},m_{1}}}} \ .
\end{equation}
The two-point function in the limit theory then becomes
\begin{align}
\langle \Phi_{q_{1},n_{1}} (z_{1},\bar{z}_{1}) \Phi_{q_{2},n_{2}} (z_{2},\bar{z}_{2})\rangle
&= \lim_{\epsilon \to 0}\lim_{k\to \infty} \alpha (k)^{2}\beta (k)^{2}
\langle \Phi_{q_{1},n_{1}}^{\epsilon ,k} (z_{1},\bar{z}_{1})
\Phi_{q_{2},n_{2}}^{\epsilon ,k} (z_{2},\bar{z}_{2})\rangle  \\
&= \lim_{\epsilon \to 0}\lim_{k\to \infty} \frac{\alpha (k)^{2}\beta
(k)^{2}}{\epsilon^{2} (k+2)^{2}}
\sum_{m \in N (q_{1},\epsilon ,k)\cap N (-q_{2},\epsilon ,k)}
 \frac{\delta_{n_{1},n_{2}}}{|z_{1}-z_{2}|^{4h_{|m|+2n_{1},m}}}\ .
\end{align}
The conformal weight $h_{|m|+2n_{1},m}$ approaches $h_{n_{1}} (q_{1}) =
(2n_{1}+1) |q_{1}|/2$ in the limit, and the sum over $m$ can be
replaced by the cardinality of the overlap,
\begin{equation}
|N (q_{1},\epsilon ,k)\cap N (-q_{2},\epsilon ,k)| = (k+2) (\epsilon
-|q_{1}+q_{2}|) \theta (\epsilon -|q_{1}+q_{2}|) + \mathcal{O} (1) \ ,
\end{equation}
where $\theta (x)$ is the Heaviside function being $1$ for positive
$x$, and $0$ otherwise. In the limit $\epsilon \to 0$ we obtain a
$\delta$-distribution,
\begin{equation}\label{delta}
\frac{\epsilon -|x|}{\epsilon^{2}}\, \theta (\epsilon -|x|)\, \to\, \delta(x)
\ . 
\end{equation}
With the choice 
\begin{equation}\label{alphabeta}
\alpha (k)\beta (k)= \sqrt{k+2}
\end{equation}
to absorb the
$k$-dependent pre-factor, we find the two-point function in a standard normalisation,
\begin{equation}
\langle \Phi_{q_{1},n_{1}} (z_{1},\bar{z}_{1}) \Phi_{q_{2},n_{2}} (z_{2},\bar{z}_{2})\rangle
= \delta_{n_{1},n_{2}} \delta (q_{1}+q_{2})
\frac{1}{|z_{1}-z_{2}|^{4h_{n_{1}} (q_{1})}} \ .
\end{equation} 
\medskip

Let us conclude by briefly discussing the limit procedure. One might
worry that the outcome depends on the precise $\epsilon$-prescription
of the limits of correlators. A conceptually clearer procedure would
be to directly define correlators of smeared fields,
\begin{equation}
\Phi_{n}[f] (z,\bar{z}) = \int dq\, f (q) \Phi_{q,n} (z,\bar{z}) \ ,
\end{equation}
by the prescription
\begin{multline}
\langle \Phi_{n_{1}}[f_{1}] (z_{1},\bar{z}_{1}) \dotsb 
\Phi_{n_{r}}[f_{r}] (z_{r},\bar{z}_{r})\rangle
= \lim_{k\to \infty} 
\beta (k)^{2} \left(\frac{\alpha (k)}{k+2} \right)^{r}\sum_{\{m_{i} \}} f_{1}
\big(-\tfrac{m_{1}}{k+2}\big)\dotsb f_{r} \big(-\tfrac{m_{r}}{k+2}\big)\\
\times \langle \phi_{|m_{1}|+2n_{1},m_{1}}(z_{1},\bar{z}_{1})\dotsb 
\phi_{|m_{r}|+2n_{r},m_{r}} (z_{r},\bar{z}_{r})\rangle \ .
\end{multline}
In this framework one would recover the correlators of the fields
$\Phi_{q,n}$ by letting the test functions $f_{i}$ approach delta
functions. Our prescription in~\eqref{defofcorrelators} corresponds to
a special choice for a family of test functions, 
\begin{equation}
f_{i} (q) = \frac{1}{\epsilon} \theta (\epsilon /2-|q-q_{i}|) \ ,
\end{equation}
which approach $\delta (q-q_{i})$ in the limit $\epsilon \to 0$, but
the result does not depend on this choice.

\section{Three-point functions}

In addition to the spectrum the three-point functions constitute the
fundamental data of a conformal field theory. In an $N=2$ superconformal theory,
all three-point functions can be derived from the correlators of three
(super-)primary fields together with the correlators involving two primaries and one
superdescendant field~\cite{Mussardo:1988av,Kiritsis:1987np}. In this section we will
analyse the limit of these correlators, which will also fix
the normalisation factors $\alpha (k)$ and $\beta(k)$. 

\subsection{Correlators of primary fields}

The correlators of three primary fields in minimal models have been
determined in~\cite{Mussardo:1988av} (they are closely related to the
three-point functions of the $SU (2)$ Wess-Zumino-Witten model derived
in~\cite{Zamolodchikov:1986bd,Dotsenko:1990zb}). Similar methods allow
the computation of correlators involving superdescendants (see
appendix~\ref{app:oddchannelthreepointfunctions}) that we will discuss
later. The correlator of three primary fields in the Neveu-Schwarz
sector in a model with diagonal spectrum reads~\cite{Mussardo:1988av}
\begin{multline}
\langle\phi_{l_{1},m_{1}} (z_{1},\bar{z}_{1}) \phi_{l_{2},m_{2}}
(z_{2},\bar{z}_{2})\phi_{l_{3},m_{3}} (z_{3},\bar{z}_{3})\rangle\\
=C(\{l_{i},m_{i}\})\delta_{m_{1}+m_{2}+m_{3},0}|z_{12}|^{2(h_{3}-h_{1}-h_{2})}|z_{13}|^{2(h_{2}-h_{1}-h_{3})}|z_{23}|^{2(h_{1}-h_{2}-h_{3})}
\end{multline}
with 
\begin{equation}
C (\{l_{i},m_{i}\})=\begin{pmatrix}
\frac{l_{1}}{2} & \frac{l_{2}}{2} & \frac{l_{3}}{2}\\
\frac{m_{1}}{2} & \frac{m_{2}}{2}& \frac{m_{3}}{2}
\end{pmatrix}^{\!2}
\sqrt{(l_{1}+1)(l_{2}+1)(l_{3}+1)}\, d_{l_{1},l_{2},l_{3}}\ .
\label{3ptcoefficient}
\end{equation}
Here, $\begin{pmatrix}
j_{1}&j_{2}&j_{3} \\
\mu_{1}&\mu_{2}&\mu_{3}
\end{pmatrix}$ denotes the Wigner 3j-symbols, and
$d_{l_{1},l_{2},l_{3}}$ is a product of Gamma functions,
\begin{equation}\label{defofd}
d_{l_{1},l_{2},l_{3}}^{2}
=\frac{\Gamma(1+\rho)}{\Gamma(1-\rho)}P^{2}(\tfrac{l_{1}+l_{2}+l_{3}+2}{2})
\prod_{k=1}^{3} \frac{\Gamma(1-\rho(l_{k}+1))}{\Gamma(1+\rho(l_{k}+1))}
\frac{P^{2}(\frac{l_{1}+l_{2}+l_{3}-2l_{k}}{2})}{P^{2}(l_{k})}
\end{equation}
with
\begin{equation}\label{def_P}
\rho=\frac{1}{k+2}\quad ,\quad
P(l)=\prod_{j=1}^{l}\frac{\Gamma(1+j\rho)}{\Gamma(1-j\rho)} \ .
\end{equation}

We want to understand the limit\footnote{In~\cite{D'Appollonio:2003dr}
a related limit of WZW models $SU (2)_{k}$ has been considered.} of
this expression when $k\to \infty$ while the labels $l_{i}$ and
$m_{i}$ grow such that the conformal weight $h$ and the $U (1)$ charge
$q$ stay constant. In particular we have
\begin{equation}
l_{i} = |m_{i}| +2n_{i}\quad \text{and} \quad m_{i} = -q_{(m_{i})}
(k+2)\ ,
\end{equation}
where $n_{i}$ is a fixed integer, and $q_{(m_{i})}$ lies in an
$\epsilon$-interval around $q_{i}$, hence it stays
approximately constant in the limit.

The Wigner 3j-symbols enforce the condition $m_{1}+m_{2}+m_{3}=0$ as
well as $l_{i_{1}}\leq l_{i_{2}}+l_{i_{3}}$ for any permutation
$\{i_{1},i_{2},i_{3} \}$ of $\{1,2,3 \}$. For definiteness we assume
now that
\begin{equation}
m_{1},m_{2}\geq 0 \quad ,\quad m_{3}=-m_{1}-m_{2} \leq 0\ .
\end{equation}
For large $|m_{i}|$ the conditions on the $l_{i}$ translate into
a single condition on the $n_{i}$,
\begin{equation}
l_{3}\leq l_{1}+l_{2} \Rightarrow n_{3}\leq n_{1}+n_{2} \ .
\end{equation}
When we consider the asymptotic behaviour of the three-point
coefficient~\eqref{3ptcoefficient} for large $k$, there are two parts
which have to be treated carefully. One is the Wigner 3j-symbol
whose limit will be discussed in appendix~\ref{app:3jsymbols}. The other
is the limit of the products of Gamma functions, where $P (l)$ becomes
an infinite product when $l$ goes to infinity. However, the infinite
products in the numerator and denominator cancel and leave a finite
product in the limit as we will show in the following.

Firstly we look at the following ratio of products of Gamma functions, 
\begin{align}
\frac{P (\frac{l_{1}+l_{2}+l_{3}+2}{2})}{P (l_{3})} &=
\frac{\prod_{j=1}^{m_{1}+m_{2}+n_{1}+n_{2}+n_{3}+1} \frac{\Gamma
(1+j\rho)}{\Gamma
(1-j\rho)}}{\prod_{j=1}^{m_{1}+m_{2}+2n_{3}}\frac{\Gamma
(1+j\rho)}{\Gamma (1-j\rho)}}\\
&= \prod_{j=m_{1}+m_{2}+2n_{3}+1}^{m_{1}+m_{2}+n_{1}+n_{2}+n_{3}+1}
\frac{\Gamma (1+j\rho)}{\Gamma (1-j\rho)} \\
&= \left(\frac{\Gamma (1+q_{(m_{3})})}{\Gamma (1-q_{(m_{3})})}
\right)^{\!n_{1}+n_{2}-n_{3}+1} \left(1+\mathcal{O} (\tfrac{1}{k}) \right)\ .
\end{align}
Similarly we have
\begin{align}
\frac{P (\frac{-l_{1}+l_{2}+l_{3}}{2})}{P (l_{2})} &=
\left(\frac{\Gamma (1+q_{(m_{2})})}{\Gamma (1-q_{(m_{2})})} 
\right)^{\!n_{1}+n_{2}-n_{3}} \left(1+\mathcal{O} (\tfrac{1}{k}) \right)\\
\frac{P (\frac{l_{1}-l_{2}+l_{3}}{2})}{P (l_{1})} &=
\left(\frac{\Gamma (1+q_{(m_{1})})}{\Gamma (1-q_{(m_{1})})}
\right)^{\!n_{1}+n_{2}-n_{3}}
\left(1+\mathcal{O} (\tfrac{1}{k}) \right)
\end{align}
and
\begin{equation}
P (\tfrac{l_{1}+l_{2}-l_{3}}{2}) = 1 +\mathcal{O} (\tfrac{1}{k})\ .
\end{equation}
In total, the coefficient $d_{l_{1},l_{2},l_{3}}$ behaves in the
limit\footnote{Note that this result can also be obtained by using the
asymptotic formula for $P$ given in~\eqref{Pasymptotics}.}
as
\begin{equation}
d_{l_{1},l_{2},l_{3}} = \left(\prod_{j=1}^{3}\frac{\Gamma
(1+q_{(m_{j})})}{\Gamma (1-q_{(m_{j})})}
\right)^{\!-\frac{1}{2}\sum_{i=1}^{3}\sigma_{i} (2n_{i}+1)}\left(1+\mathcal{O} (\tfrac{1}{k}) \right) \ .
\end{equation}
Here, $\sigma_{i}=\text{sgn} (q_{i})$ denotes the sign of the
corresponding charge. In this form the expression is valid without any
assumptions on which of the charges are positive or negative.

The asymptotic behaviour of the 3j-symbols is derived in
appendix~\ref{app:3jsymbols}. For $m_{i}$ linearly growing with $k$ and
$m_{1},m_{2}\geq 0$, $m_{3}\leq 0$, it is given by
(see~\eqref{app:3jasymptotic})
\begin{equation}
\begin{pmatrix}
\frac{|m_{1}|}{2}+n_{1} & \frac{|m_{2}|}{2}+n_{2} & \frac{|m_{3}|}{2}+n_{3}\\
\frac{m_{1}}{2} & \frac{m_{2}}{2} & \frac{m_{3}}{2}
\end{pmatrix} = (-1)^{m_{1}+n_{3}+n_{2}} \left(|m_{3}| \right)^{-1/2}
d^{J}_{M',M}
(\beta) \cdot (1+\mathcal{O} (\tfrac{1}{k}))\ ,
\end{equation}
where $d^{J}_{M',M} (\beta)$ is the Wigner $d$-matrix and 
\begin{equation}\label{betaJMM}
\cos \beta
=\frac{|m_{1}|-|m_{2}|}{|m_{1}|+|m_{2}|}\ , \
J=\frac{n_{1}+n_{2}}{2}\ ,\ M' =
-\frac{n_{1}+n_{2}}{2}+n_{3}\ ,\ M=\frac{n_{1}+n_{2}}{2}-n_{2}
\ .
\end{equation}

Putting everything together, the three-point coefficient 
$C(\{l_{i},m_{i}\})$ given in~\eqref{3ptcoefficient} has the limiting
behaviour
\begin{equation}
C (\{l_{i},m_{i}\}) \sim (k+2)^{1/2} \mathcal{C} (\{q_{(m_{i})},n_{i}\})\ ,
\end{equation}
where $\mathcal{C}$ is a smooth function of the charges $q_{i}$. For
$q_{1},q_{2}<0$ and $q_{3}>0$ it is given by
\begin{equation}\label{Ccal}
\mathcal{C} (\{q_{i},n_{i} \}) 
= \left( \frac{|q_{1}q_{2}|}{|q_{3}|}\right)^{1/2}
(d^{J}_{M',M}(\beta))^{2} 
\left(\prod_{j=1}^{3}\frac{\Gamma
(1+q_{j})}{\Gamma (1-q_{j})}
\right)^{\!n_{1}+n_{2}-n_{3}+\frac{1}{2}} \ ,
\end{equation}
with $\cos \beta = \frac{|q_{1}|-|q_{2}|}{|q_{1}|+|q_{2}|}$ and
$J,M,M'$ given in~\eqref{betaJMM}. Notice that $\mathcal{C}$ in this case is non-zero
only for $n_{1}+n_{2}\geq n_{3}$.

\begin{figure}
\begin{center}
\includegraphics{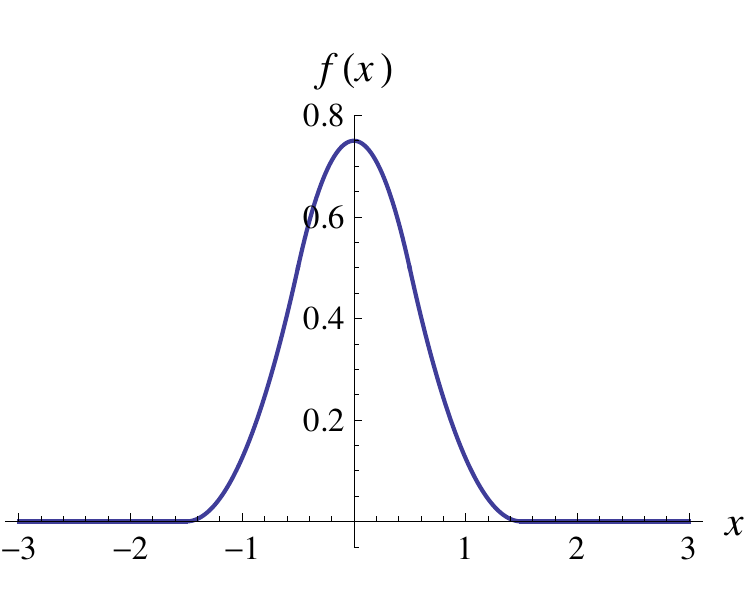}  
\end{center}
\caption{\label{fig:fPlot}An illustration of the function $f$ defined in~\eqref{def_f}.}
\end{figure}

Now we are ready to work out the limit of the 3-point function. By
definition it is given by 
\begin{align}
& \langle \Phi_{q_{1},n_{1}} (z_{1},\bar{z}_{1}) \Phi_{q_{2},n_{2}} (z_{2},\bar{z}_{2}) \Phi_{q_{3},n_{3}} (z_{3},\bar{z}_{3})\rangle\nonumber\\
&\qquad 
= \lim_{\epsilon \to 0} \lim_{k\to \infty} 
\beta (k)^{2} \alpha (k)^{3} \langle \Phi_{q_{1},n_{1}}^{\epsilon ,k}
(z_{1},\bar{z}_{1})\Phi_{q_{2},n_{2}}^{\epsilon ,k} (z_{2},\bar{z}_{2})
\Phi_{q_{3},n_{3}}^{\epsilon ,k} (z_{3},\bar{z}_{3})\rangle \\
& \qquad = \lim_{\epsilon \to 0} \lim_{k\to \infty} \frac{\beta (k)^{2}
\alpha (k)^{3}}{\epsilon^{3} (k+2)^{3}} \sum_{\{ m_{i}\in N
(q_{i},\epsilon ,k)\}} C (\{|m_{i}|+n_{i},m_{i} \})
\delta_{m_{1}+m_{2}+m_{3},0}\nonumber\\
&\qquad \qquad \times 
|z_{12}|^{2(h_{3}-h_{1}-h_{2})}|z_{13}|^{2(h_{2}-h_{1}-h_{3})}|z_{23}|^{2(h_{1}-h_{2}-h_{3})} 
\ .
\end{align}
We already determined the limit of the three-point coefficient, so it
only remains to determine the factor that originates from the summation
over the labels $m_{i}$, i.e.\ the cardinality of the set
\begin{equation}
N_{123}= \{(m_{1},m_{2},m_{3}) \in N (q_{1},\epsilon ,k) \times N (q_{2},\epsilon ,k) \times N (q_{3},\epsilon ,k):m_{1}+m_{2}+m_{3}=0
\} \ .
\end{equation}
It is given by
\begin{equation}
|N_{123}| = (k+2)^{2}\epsilon^{2} f ({\textstyle \frac{1}{\epsilon}\sum_{i}q_{i}}) +\mathcal{O} (k+2) \ ,
\end{equation}
where the function $f$ is defined as
\begin{equation}
f (x) = \left\{\begin{array}{crr@{\,x\,}l}
0 & \quad \text{for} & & < -\frac{3}{2} \\[1mm]
\frac{1}{2} (x+\frac{3}{2})^{2}& \quad \text{for} & -\frac{3}{2}< & < -\frac{1}{2} \\[1mm]
\frac{3}{4}-x^{2}& \quad \text{for} & -\frac{1}{2}< & < \frac{1}{2} \\[1mm]
\frac{1}{2} (x-\frac{3}{2})^{2}& \quad \text{for} & \frac{1}{2}< & < \frac{3}{2} \\[1mm]
 0 & \quad \text{for} & \frac{3}{2}< & \ .
\end{array} \right. 
\label{def_f}
\end{equation}
The function $f$ is displayed in figure~\ref{fig:fPlot}, it has the 
property
\begin{equation}
\int dx\, f (x) = 1 \ .
\end{equation}

When we finally take the limit, we observe that the function $f$ leads
to a delta distribution for the sum of the charges,
\begin{equation}
\frac{1}{\epsilon}f ({\textstyle \frac{1}{\epsilon }\sum_{i}q_{i}}) \to \delta
({\textstyle\sum_{i}q_{i}}) \ .
\end{equation}
Using the condition~\eqref{alphabeta} we can absorb the remaining
$k$-dependence by setting
\begin{equation}
\alpha (k) = (k+2)^{-1/2} \quad ,\quad \beta (k) = (k+2)  \ .
\end{equation}
The total result is then
\begin{multline}
\langle \Phi_{q_{1},n_{1}} (z_{1},\bar{z}_{1}) \Phi_{q_{2},n_{2}} (z_{2},\bar{z}_{2}) \Phi_{q_{3},n_{3}} (z_{3},\bar{z}_{3})\rangle \\
= \mathcal{C} (\{q_{i},n_{i} \}) \delta ({\textstyle \sum_{i}q_{i}})
|z_{12}|^{2(h_{3}-h_{1}-h_{2})}|z_{13}|^{2(h_{2}-h_{1}-h_{3})}|z_{23}|^{2(h_{1}-h_{2}-h_{3})} 
\end{multline}
with $\mathcal{C}$ given in~\eqref{Ccal}. 

\subsection{Correlators involving superdescendants}

Now we want to show that also the three-point function of two
primaries and one superdescendant (which corresponds to the odd fusion
channel~\cite{Mussardo:1988av}) has a well-defined limit. We will
limit ourselves to the case of a superdescendant obtained by acting
with $G^{+}$, the discussion for $G^{-}$-descendants is analogous. As derived
in appendix~\ref{app:oddchannelthreepointfunctions} such a correlator is given by
(see~\eqref{app:oddcorrelatorresult})
\begin{align}
& \langle (\bar{G}^{+}_{-\frac{1}{2}}G^{+}_{-\frac{1}{2}}\phi_{l_{1},m_{1}}) (z_{1},\bar{z}_{1})
\phi_{l_{2},m_{2}} (z_{2},\bar{z}_{2})\phi_{l_{3},m_{3}}
(z_{3},\bar{z}_{3}) \rangle \nonumber\\
&\quad = \frac{k+2}{2 (n_{1}+1) (l_{1}-n_{1})}
\begin{pmatrix} \frac{l_{2}+m_{2}}{2} \\ \frac{l_{1}-m_{1}}{2}+1 \end{pmatrix}
\begin{pmatrix} \frac{l_{1}+l_{2}-m_{1}-m_{2}}{2}+1 \\ \frac{l_{1}-m_{1}}{2}+1\end{pmatrix}
\begin{pmatrix} l_{1} \\ \frac{l_{1}-m_{1}}{2}+1 \end{pmatrix}^{\!-1}\nonumber\\
&\qquad  \times \begin{pmatrix}
\frac{k-l_{1}}{2} & \frac{l_{2}}{2} & \frac{l_{3}}{2} \\
-\frac{k-l_{1}}{2} & \frac{m_{1}+m_{2}-l_{1}}{2}-1 & \frac{m_{3}}{2}
\end{pmatrix}^{\!2}
\sqrt{(k-l_{1}+1) (l_{2}+1) (l_{3}+1)}\, 
d_{k-l_{1},l_{2},l_{3}}\nonumber\\
&\qquad  \times 
|z_{12}|^{2 (h_{l_{3},m_{3}}-(h_{l_{1},m_{1}}+1/2)-h_{l_{2},m_{2}})} 
|z_{23}|^{2 ((h_{l_{1},m_{1}}+1/2)-h_{l_{2},m_{2}}-h_{l_{3},m_{3}})}\nonumber\\
&\qquad \times 
|z_{13}|^{2 (h_{l_{2},m_{2}}- (h_{l_{1},m_{1}}+1/2)-h_{l_{3},m_{3}})}\ ,
\label{oddcorrelator}
\end{align}
where $l_{i}\geq |m_{i}|$ and we assume that $m_{1},m_{2}>0$ and
$m_{3}<0$.  

To determine the limit we first simplify the prefactor (that we call $A$)
in~\eqref{oddcorrelator} by expressing the 3j-symbol with the help
of~\eqref{app:extremal3j},
\begin{align}
A &=  \frac{k+2}{2 (n_{1}+1) (l_{1}-n_{1})}
\begin{pmatrix} \frac{l_{2}+m_{2}}{2} \\ \frac{l_{1}-m_{1}}{2}+1 \end{pmatrix}
\begin{pmatrix} \frac{l_{1}+l_{2}-m_{1}-m_{2}}{2}+1 \\ \frac{l_{1}-m_{1}}{2}+1\end{pmatrix}
\begin{pmatrix} l_{1} \\ \frac{l_{1}-m_{1}}{2}+1 \end{pmatrix}^{\!-1}\nonumber\\
&\qquad  \times \begin{pmatrix}
\frac{k-l_{1}}{2} & \frac{l_{2}}{2} & \frac{l_{3}}{2} \\
-\frac{k-l_{1}}{2} & \frac{m_{1}+m_{2}-l_{1}}{2}-1 & \frac{m_{3}}{2}
\end{pmatrix}^{\!2} \nonumber\\
& = \frac{k+2}{2 (n_{1}+1) (l_{1}-n_{1})}\, \frac{(\frac{l_{1}+m_{1}}{2}-1)!}{l_{1}!(\frac{l_{1}-m_{1}}{2}+1)!} 
\nonumber\\
&\qquad \times \frac{(\frac{-k+l_{1}+l_{2}+l_{3}}{2})!(\frac{l_{3}+m_{3}}{2})!(\frac{l_{2}+m_{2}}{2})!(k-l_{1})!}{(\frac{k-l_{1}+l_{2}+l_{3}}{2}+1)!(\frac{k-l_{1}-l_{2}+l_{3}}{2})!(\frac{k-l_{1}+l_{2}-l_{3}}{2})!(\frac{l_{3}-m_{3}}{2})!(\frac{l_{2}-m_{2}}{2})!}
\ .
\end{align}
In the limit we set $l_{i}=|m_{i}|+2n_{i}$ where the $n_{i}$ are kept
constant, and the $m_{i}$ are sent to infinity growing linearly in
$k$. By using~\eqref{limitoffactorials} we get
\begin{align}
A & = \frac{k+2}{2 (n_{1}+1) (m_{1}+n_{1})} \,\frac{n_{3}!}{(n_{1}+1)!(n_{2})!(-n_{1}-n_{2}+n_{3}-1)!}\nonumber\\
&\quad \times \frac{(m_{1}+n_{1}-1)!(k-m_{1}-2n_{1})!(m_{2}+n_{2})!(-m_{3}+n_{1}+n_{2}+n_{3}+1)!}{(m_{1}+2n_{1})!(k-m_{1}-n_{1}+n_{2}-n_{3}+1)!(m_{2}-n_{1}+n_{2}+n_{3})!(-m_{3}+n_{3})!}\nonumber\\
& = \frac{1}{2 (k+2) (n_{1}+1)}\, \frac{n_{3}!}{ (n_{1}+1)!n_{2}!(-n_{1}-n_{2}+n_{3}-1)!}\nonumber\\
& \qquad \times
|q_{1}|^{-n_{1}-2}|1+q_{1}|^{-n_{1}-n_{2}+n_{3}-1}|q_{2}|^{n_{1}-n_{3}}|q_{3}|^{n_{1}+n_{2}+1}\left(
1+\mathcal{O} (\tfrac{1}{k})\right)\ ,
\end{align}
where $q_{i}=-\frac{m_{i}}{k+2}$ is kept fixed in the limit. By
similar arguments as before we can evaluate the asymptotic form of
$d_{k-l_{1},l_{2},l_{3}}$ to be
\begin{multline}
d_{k-l_{1},l_{2},l_{3}} =\left(\frac{\Gamma (1+|1+q_{1}|)\Gamma
(1-|q_{2}|)\Gamma (1+|q_{3}|)}{\Gamma (1-|1+q_{1}|)\Gamma
(1+|q_{2}|)\Gamma (1-|q_{3}|)}
\right)^{\!n_{1}+n_{2}-n_{3}+\frac{1}{2}} \cdot \left(1+\mathcal{O}
(\tfrac{1}{k+2}) \right)\ .
\end{multline}
The final result for the three-point correlator of two primaries and
one superdescendant in the limit theory is then given by
\begin{align}
&\langle
(G^{+}_{-\frac{1}{2}}\bar{G}^{+}_{-\frac{1}{2}}\Phi_{q_{1},n_{1}})
(z_{1},\bar{z}_{1}) \Phi_{q_{2},n_{2}}
(z_{2},\bar{z}_{2})\Phi_{q_{3},n_{3}} (z_{3},\bar{z}_{3})\rangle = \nonumber\\
&\quad = \frac{1}{2 (n_{1}+1)}\,
\frac{n_{3}!}{(n_{1}+1)!n_{2}!(n_{3}-n_{1}-n_{2}-1)!}
|1+q_{1}|^{-n_{1}+n_{2}+n_{3}-\frac{1}{2}}
|q_{2}|^{n_{1}-n_{3}+\frac{1}{2}}|q_{3}|^{n_{1}+n_{2}+\frac{3}{2}}\nonumber\\
& \qquad \times |q_{1}|^{-n_{1}-2}\left(\frac{\Gamma (1+|1+q_{1}|)\Gamma
(1-|q_{2}|)\Gamma (1+|q_{3}|)}{\Gamma (1-|1+q_{1}|)\Gamma
(1+|q_{2}|)\Gamma (1-|q_{3}|)} \right)^{\!n_{1}+n_{2}-n_{3}+\frac{1}{2}}\nonumber\\
& \qquad \times \delta (1+q_{1}+q_{2}+q_{3}) |z_{12}|^{2
(h_{3}-h_{1}-\frac{1}{2}-h_{2})}|z_{13}|^{2
(h_{2}-h_{1}-\frac{1}{2}-h_{3})}|z_{23}|^{2
(h_{1}+\frac{1}{2}-h_{2}-h_{3})} \ ,
\end{align}
where we assumed that $q_{1},q_{2}<0$ and $q_{3}>0$. The
generalisation to other cases is straightforward. As in a
superconformal theory all three-point functions are determined if
the three-point correlators of three primaries and the correlators of
two primaries and one superdescendant are given, this result shows that all
three-point functions of the limit theory are well defined.

\section{Boundary conditions and one-point functions}

In this section we investigate the limit of bulk one-point functions
on the upper half plane. In the minimal models there are two types of
maximally symmetric boundary conditions, called A-type and
B-type~\cite{Maldacena:2001ky}. In a diagonal model, only chargeless
fields can couple to B-type boundary conditions, so that we do not
find any B-type boundary conditions in the limit theory. On the other
hand we will discover two families of A-type
boundary conditions in the limit.

The A-type boundary conditions are labelled by the same labels as the
representations of the bosonic subalgebra of the superconformal
algebra, $(L,M,S)$, where $L$ is an integer satisfying $0\leq
L\leq k$, $M$ is a $(2k+4)$-periodic integer, and $S$ is a $4$-periodic
integer such that $L+M+S$ is even. Labels are identified according
to~\eqref{cosetident}.

Boundary states $|L,M,S\rangle$ are given by linear combinations of
Ishibashi states~\cite{Maldacena:2001ky},
\begin{equation}
|L,M,S\rangle  = \sum_{(l,m,s)}\frac{S_{(L,M,S)
(l,m,s)}}{\sqrt{S_{(0,0,0) (l,m,s)}}}
|l,m,s\rangle \!\rangle \ ,
\end{equation}
where $S$ is the modular S-matrix of the $N=2$ superconformal algebra,
\begin{equation}
S_{(L,M,S)(l,m,s)}=\frac{1}{k+2}\sin{\frac{\pi(l+1)(L+1)}{k+2}}e^{-\pi i (\frac{s S}{2}-\frac{m M}{k+2})}\ .
\end{equation}

The coefficients of the boundary states determine the bulk one-point
functions  on the upper half plane with boundary condition $\alpha = (L,M,S)$ on
the real axis. For a primary field $\phi_{l,m}$ it is given by
\begin{equation}
\langle \phi_{l,m} (z,\bar{z}) \rangle_{(L,M,S)} = 
\frac{S_{(L,M,S) (l,m,0)}}{\sqrt{S_{(0,0,0) (l,m,0)}}} |z-\overline{z}|^{-2 h_{l,m}}\ .
\end{equation}
Writing out the one-point function we get
\begin{equation}
\langle \phi_{l,m} (z,\bar{z}) \rangle_{(L,M,S)} = 
(k+2)^{-1/2} \frac{\sin \frac{\pi (l+1) (L+1)}{k+2}}{\sqrt{\sin
\frac{\pi (l+1)}{k+2}}} e^{\pi i \frac{mM}{k+2}} 
|z-\overline{z}|^{-2 h_{l,m}}\ . 
\end{equation}
When we take the limit $k\to \infty$ we have some freedom of what to do
with the boundary labels. There are two natural choices: either we
keep the boundary labels constant in the limit, or we scale them in
the same way as we scale the field labels. Both lead to sensible
expressions as we will see shortly.

\subsection{Discrete boundary conditions}

First we will take the limit such that the boundary labels are kept
fixed. The one-point function in the limit is then
\begin{align}
\langle \Phi_{q,n} (z,\bar{z})\rangle_{(L,M,S)} & = \lim_{\epsilon
\to 0} \lim_{k\to\infty} \alpha (k)\beta (k) \langle
\Phi_{q,n}^{\epsilon ,k} (z,\bar{z}) \rangle_{(L,M,S)} \nonumber\\
& =  \lim_{\epsilon \to 0} \lim_{k\to\infty} \frac{\alpha (k)\beta
(k)}{\epsilon (k+2)^{\frac{3}{2}}} \sum_{m\in N (q,\epsilon ,k)} 
\frac{\sin \frac{\pi (|m|+2n+1) (L+1)}{k+2}}{\sqrt{\sin
\frac{\pi (|m|+2n+1)}{k+2}}} e^{\pi i \frac{mM}{k+2}} 
|z-\overline{z}|^{-2 h_{|m|+2n,m}}\nonumber\\
& = \frac{\sin (\pi |q| (L+1))}{\sqrt{\sin (\pi |q|)}} e^{-\pi i qM} 
|z-\overline{z}|^{-2 h_{n} (q)} \ . 
\end{align}
For the Ramond fields one finds
\begin{align}
\langle \Psi^{0}_{q} (z,\bar{z}) \rangle_{(L,M,S)} &= 
 \frac{\sin (\pi |\frac{1}{2}-q| (L+1))}{\sqrt{\sin (\pi
 |\frac{1}{2}-q|)}} e^{\pi i (\frac{1}{2}-q) M}e^{-\pi i\frac{S}{2}}  
|z-\overline{z}|^{-1/4}\\
\langle \Psi^{\pm}_{q,n} (z,\bar{z}) \rangle_{(L,M,S)} &= 
 \frac{\sin (\pi |\frac{1}{2}\mp q| (L+1))}{\sqrt{\sin (\pi
 |\frac{1}{2}\mp q|)}} e^{\pi i (\pm\frac{1}{2}-q) M}e^{\mp \pi i\frac{S}{2}}  
|z-\overline{z}|^{-2 h^{\pm}_{n} (q)} \ .
\end{align}
These boundary conditions are not independent. Using the trigonometric
identity
\begin{equation}
\sin \left(\pi |q| (L+1) \right) = \sin \left(\pi |q| \right)
\sum_{j=0}^{L} e^{i\pi q (L-2j)} \ ,
\end{equation}
we see that
\begin{equation}
\langle \ \cdot\ \rangle_{(L,M,S)} = \sum_{j=0}^{L} \langle
\ \cdot\ \rangle_{(0,M+L-2j,S)} \ .
\end{equation}
All boundary conditions are therefore superpositions of boundary
conditions with $L=0$, and the elementary boundary conditions are
$(0,M,S)$. This can be compared to the situation in minimal models
before taking the limit, where all boundary conditions can be obtained
by boundary renormalisation group flows from superpositions of those
with $L=0$~\cite{Fredenhagen:2001nc,Maldacena:2001ky}. These flows
become shorter when the level $k$ grows, and in the limit the boundary
conditions can be identified.

\subsection{Continuous boundary conditions}

Now we will scale the boundary labels in the same way as we did for
the field labels. We introduce a continuous parameter $Q$, $0<|Q|<1$,
and a discrete parameter $N\in\mathbb{N}_{0}$, and instead of
considering fixed boundary labels in the limit, we consider a sequence
of boundary conditions $B_{k} (Q,N)$ of the form
\begin{equation}
B_{k}(Q,N) = (|\lfloor -Q (k+2) \rfloor|+2N,\lfloor -Q
(k+2)\rfloor,0) \ ,
\end{equation}
where $\lfloor x \rfloor$ denotes the greatest integer smaller or
equal to $x$.
The one-point function in the limit is then
\begin{align}
\langle \Phi_{q,n} (z,\bar{z})\rangle_{(Q,N)} & = \lim_{\epsilon
\to 0} \lim_{k\to\infty} \alpha (k)\beta (k) \langle
\Phi_{q,n}^{\epsilon ,k} (z,\bar{z}) \rangle_{B_{k} (Q,N)} \\
& =  \lim_{\epsilon \to 0} \lim_{k\to\infty} \frac{\alpha (k)\beta
(k)}{\epsilon (k+2)^{\frac{3}{2}}} \sum_{m\in N (q,\epsilon ,k)} 
\frac{\sin \frac{\pi (|m|+2n+1) (|\lfloor -Q (k+2)\rfloor|+2N+1)}{k+2}}{\sqrt{\sin
\frac{\pi (|m|+2n+1)}{k+2}}} \nonumber \\
&\qquad \times e^{\pi i \frac{m \lfloor -Q (k+2)\rfloor}{k+2}} 
|z-\overline{z}|^{-2 h_{|m|+2n,m}} \ .
\end{align}
We observe that the arguments of the sine function in the numerator
and of the exponential diverge when $k$ is sent to infinity, so that
we get strongly oscillating expressions. Their combination behaves as
\begin{align}
& 2i\sin \frac{\pi (|m|+2n+1) (|\lfloor -Q (k+2)\rfloor|+2N+1)}{k+2} 
e^{\pi i \frac{m \lfloor -Q (k+2)\rfloor}{k+2}}\nonumber\\
& \sim\left(e^{i\frac{\pi |m||\lfloor -Q
(k+2)\rfloor|}{k+2}}e^{i\frac{\pi [|m| (2N+1)+ (2n+1)|\lfloor -Q
(k+2)\rfloor|]}{k+2}} - e^{-i\frac{\pi |m||\lfloor -Q
(k+2)\rfloor|}{k+2}}e^{-i\frac{\pi [|m| (2N+1)+ (2n+1)|\lfloor -Q
(k+2)\rfloor|]}{k+2}}\right) \nonumber \\
& \quad \times e^{i \frac{\pi m \lfloor -Q (k+2)\rfloor}{k+2}}\\
& \sim \left\{ \begin{array}{ll}
\left(e^{2i\frac{\pi |m||\lfloor -Q
(k+2)\rfloor|}{k+2}}e^{i\pi [|q| (2N+1)+ (2n+1)|Q|]} - e^{-i\pi [|q| (2N+1)+ (2n+1)|Q|]}\right) & \text{for} \ qQ>0\\[3mm]
\left(e^{i\pi [|q| (2N+1)+ (2n+1)|Q|]} - e^{-2i\frac{\pi |m||\lfloor -Q
(k+2)\rfloor|}{k+2}}e^{-i\pi [|q| (2N+1)+ (2n+1)|Q|]}\right) & \text{for} \ qQ<0 \ .
\end{array} \right. 
\end{align}
Upon taking the average over $m$ the strongly oscillating term is
suppressed, and in the limit only the other term survives. The final
result is therefore
\begin{equation}
\langle \Phi_{q,n} (z,\bar{z})\rangle_{(Q,N)} = 
\frac{1}{2i\sqrt{\sin (\pi |q|)}}|z-\bar{z}|^{-2 h_{n} (q)}\times \left\{\begin{array}{ll}
- e^{-i\pi [|q| (2N+1)+ (2n+1)|Q|]} & \text{for} \ qQ>0\\[2mm]
e^{i\pi [|q| (2N+1)+ (2n+1)|Q|]} & \text{for} \ qQ<0\ .
\end{array} \right.
\end{equation}
Similarly, in the Ramond sector we find
\begin{align}
\langle \Psi^{0}_{q} (z,\bar{z})\rangle_{(Q,N)} &= 
\frac{e^{-\pi i\frac{S}{2}}}{2i\sqrt{\sin (\pi |\frac{1}{2}-q|)}}|z-\bar{z}|^{-1/4}\times \left\{\begin{array}{ll}
- e^{-i\pi|\frac{1}{2}-q| (2N+1)} & \text{for} \ Q>0\\[2mm]
e^{i\pi |\frac{1}{2}-q| (2N+1)} & \text{for} \ Q<0 
\end{array} \right. \\
\langle \Psi^{\pm}_{q,n} (z,\bar{z})\rangle_{(Q,N)} &= 
\frac{e^{\mp \pi i\frac{S}{2}}|z-\bar{z}|^{-2 h^{\pm}_{n} (q)}}{2i\sqrt{\sin (\pi |q\mp\frac{1}{2}|)}}\times \left\{\begin{array}{ll}
- e^{-i\pi [|q\mp \frac{1}{2}| (2N+1)+ 2n|Q|]} & \text{for} \ (q\mp
\frac{1}{2})Q>0\\[2mm]
e^{i\pi [|q\mp \frac{1}{2}| (2N+1)+ 2n|Q|]} & \text{for} \ (q\mp
\frac{1}{2}) Q<0\ .
\end{array} \right.
\end{align}

\section{Fields of charge zero}
\label{sec:chargezero}
When we consider the limit of the spectrum (see
figure~\ref{fig:spectrum}), we observe that there
are also fields of charge zero in the minimal model, which do not
contribute to the averaged fields~$\Phi_{q,n}$. This raises the
question whether we can define another class of fields in the limit
theory that arises from chargeless fields in the minimal model.

Following our general strategy of defining fields in the limit theory
leads to an ansatz for chargeless fields which does not seem to give a
sensible result. Therefore we will follow a different ansatz in
section~\ref{sec:secondansatz}. The latter one appears to give a
sensible class of fields which however is decoupled from the fields
$\Phi_{q,n}$. 

\subsection{First ansatz: average of approximately chargeless fields}

In the spirit of our general construction, we should try to define
possible chargeless fields by averaging over fields that approximate a
given conformal weight $h$ and a $U (1)$ charge $q=0$ in the limit.
In contrast to our analysis in section~\ref{sec:spectrum} the label
$m$ now has to stay small compared to $k$, so that we do not need a
strong fine-tuning in the growth of $l$ and $m$ to get a finite weight
$h$ (recall that the difference $l-|m|=2n$ stays finite in that
case). Instead both terms in the formula~\eqref{def_hlm} for the weight
$h_{l,m,0}$, the one depending on $l$ and the one depending on $m$
contribute to the weight on an equal footing. Therefore we are led to
introduce labels $y,p,\mu $ such that
\begin{equation}
h=\frac{y^{2}}{4} \quad ,\quad l=p\sqrt{k+2}\quad ,\quad m=v
\sqrt{k+2} \ .
\end{equation} 
For large quantum numbers $l$ we then get the relation
\begin{equation}
y^{2} = p^{2}-v^{2}\ .
\end{equation}
The condition that the charge $q=-\frac{m}{k+2}$ is close to zero,
$|q|<\epsilon /2$, translates into a condition on $v$,
\begin{equation}
|v |<\frac{1}{2}\epsilon \sqrt{k+2} \ .
\end{equation}
We define the set of labels $\hat{N} (y,\delta ,\epsilon ,k)$
that correspond to fields with charge approximately zero, and
conformal weight close to $y^{2}/4$,
\begin{align}
\hat{N} (y,\delta ,\epsilon ,k) &= \big\{ (l,m)| |m|\leq l,\
|m|<\tfrac{\epsilon}{2}
(k+2),\ |y-2\sqrt{h_{l,m}}|<\tfrac{\delta}{2} \big\} \\
&=  \big\{ (l,m)| |m|\leq l,\ |v |<\tfrac{\epsilon}{2}
\sqrt{k+2},\ |y-\sqrt{p^{2}-v^{2}}|<\tfrac{\delta}{2} \big\}\ .
\end{align}
The cardinality of $\hat{N}$ for large level is given by
\begin{equation}
|\hat{N} (y,\delta ,\epsilon ,k)| = (k+2) y\delta \log
\frac{\epsilon^{2} (k+2)}{y^{2}} \left(1+\mathcal{O} (\delta)\right)
+\mathcal{O} (\log (k+2))\ ,
\end{equation}
where in the leading term in $(k+2)$ we only stated the linear term in
$\delta$.

This suggests to define the averaged fields
\begin{equation}
\hat{\Phi}_{y}^{\delta ,\epsilon ,k} := \frac{1}{|\hat{N} (y,\delta
,\epsilon ,k)|} \sum_{(l,m)\in \hat{N} (y,\delta
,\epsilon ,k)} \phi_{l,m} \ .
\end{equation}
Introducing a normalisation factor $\hat{\alpha} (y,k)$, we obtain the
two-point function for the limit fields $\hat{\Phi}_{y}$,
\begin{align}
&\langle \hat{\Phi}_{y_{1}} (z_{1},\bar{z}_{1})\hat{\Phi}_{y_{2}}
(z_{2},\bar{z}_{2})\rangle \nonumber\\
&\quad = \lim_{\epsilon ,\delta \to 0}
\lim_{k\to \infty}
\frac{\hat{\alpha} (y_{1},k)\hat{\alpha} (y_{2},k)\beta(k)^{2}}{|\hat{N} (y_{1},\delta
,\epsilon ,k)||\hat{N} (y_{2},\delta
,\epsilon ,k)|} \sum_{(l_{i},m_{i})\in\hat{N} (y_{i},\delta
,\epsilon ,k) }\frac{\delta_{l_{1},l_{2}}\delta_{m_{1},-m_{2}}}{|z_{1}-z_{2}|^{4h_{l_{1},m_{1}}}}
\\
& \quad = \frac{\delta (y_{1}-y_{2})}{|z_{1}-z_{2}|^{y_{1}^{2}}}
\ ,
\end{align}
for $\hat{\alpha} (y,k)= \sqrt{y\frac{\log (k+2)}{k+2}}$. We observe
that in contrast to the analysis in section~\ref{sec:spectrum} the
normalisation factor $\hat{\alpha}$ now depends on the field label.

This is only the first oddity of this construction. The main problem
of this ansatz is that the fields $\hat{\Phi}_{y}$ come out as an
average over fields with different values of $p$ and $v$, whereas
the correlators heavily depend on $p$ and $v$, so that the fields
over which we average do not tend to have a similar behaviour in the
limit. This is against the spirit of the limiting procedure because we
only want to combine fields that have similar behaviour. One can now
explicitly check that correlators involving these fields do not have a
well-defined limit. For example one can easily check that the
one-point function of $\hat{\Phi}_{y}$ diverges for the $(0,0,0)$
boundary condition.

All in all, these results suggest that our ansatz for $\hat{\Phi}$
does not lead to a sensible field in the limit theory. On the other
hand, the failure mainly resulted from the attempt to average over
fields that do not behave similarly in the limit. Instead we will now
try to define fields where we keep the quantum number $m=0$ fixed in
the limit. These fields tend to behave much better in the limit,
although they decouple from the charged fields $\Phi_{q,n}$. Their
behaviour seems to point towards the existence of another consistent
limit theory that is decoupled from the one that we discussed
before. This will be further explored in~\cite{Fredenhagen:inpreparation}.

\subsection{Second ansatz starting from exactly chargeless fields}
\label{sec:secondansatz}
Again we introduce a label $p>0$ such that $h=\frac{p^{2}}{4}$. For
fields $\phi_{l,0}$ that approach a conformal weight $h$ the
label $l$ behaves as
\begin{equation}\label{l_forchargezero}
l = p\sqrt{k+2} -1 +\mathcal{O} (1/k^{1/2})\ ,
\end{equation}
it grows with the square root of $k$.  In addition to the spectrum
concentrated on lines of slope $(2n+1)$ that we found in
section~\ref{sec:spectrum}, we thus can try to define another
continuous class of fields $\tilde{\Phi}_{p}$ that have $U(1)$-charge
$0$ and conformal weight $h (p) =p^{2}/4$.

Similarly we set up fields $\tilde{\Psi}^{\pm}_{p}$ in the Ramond
sector with $q=\pm \frac{1}{2}$ and
$h=\frac{1}{8}+\frac{p^{2}}{4}$. They arise from fields
$\psi^{\pm}_{l,0}$ with $l\approx p\sqrt{k+2}$.

We now want to analyse correlators of the fields $\tilde{\Phi}_{p}$ of
zero charge in the limit theory. As we have seen
in~\eqref{l_forchargezero}, those fields have to be defined in terms
of minimal model fields $\phi_{l,0}$, where the label $l$ grows with
the square root of $k+2$. 
We introduce the averaged fields
\begin{equation}
\tilde{\Phi}^{\epsilon ,k}_{p} = \frac{1}{\left|\tilde{N} (p,\epsilon
,k)\right|} 
\sum_{l\in \tilde{N} (p,\epsilon ,k)} \phi_{l,0} \ .
\end{equation}
The set $\tilde{N} (p,\epsilon ,k)$ contains those labels $l$ such
that the corresponding conformal weight is close to $h (p)= p^{2}/4$,
\begin{equation}
\tilde{N} (p,\epsilon ,k) = \left\{ l: p-\frac{\epsilon}{2} <
\frac{l}{\sqrt{k+2}} < p + \frac{\epsilon}{2}  \right\} \ .
\end{equation}
For large $k$ its cardinality is
\begin{equation}
\left|\tilde{N} (p,\epsilon ,k)\right| = \epsilon \sqrt{k+2} + \mathcal{O} (1) \ .
\end{equation}
Here we assumed $p>0$ and $\epsilon$ small enough such that
$p-\frac{\epsilon}{2}>0$.

The two-point function of such fields in the limit theory is then
\begin{align}
\langle \tilde{\Phi}_{p_{1}} (z_{1},\bar{z}_{1}) \tilde{\Phi}_{p_{2}} (z_{2},\bar{z}_{2})\rangle
& = \lim_{\epsilon \to 0}\lim_{k\to \infty} \tilde{\alpha} (k)^{2}\beta
(k)^{2} \langle \tilde{\Phi}^{\epsilon ,k}_{p_{1}} (z_{1},\bar{z}_{1})
\tilde{\Phi}_{p_{2}}^{\epsilon ,k} (z_{2},\bar{z}_{2})\rangle\\
& =  \lim_{\epsilon \to 0}\lim_{k\to \infty} \frac{\tilde{\alpha} (k)^{2}\beta
(k)^{2}}{\epsilon^{2} (k+2)} \sum_{l\in \tilde{N} (p_{1},\epsilon
,k)\cap \tilde{N} (p_{2},\epsilon ,k)} \frac{1}{|z_{1}-z_{2}|^{4h_{l,0}}}
\ .
\end{align}
We introduced a new normalisation factor $\tilde{\alpha} (k)$ for
the charge zero fields.
The conformal weight $h_{l,0}$ approaches $h (p_{1})=p_{1}^{2}/4$, and
the sum can be replaced by the cardinality of the overlap,
\begin{equation}
\left|\tilde{N} (p_{1},\epsilon ,k)\cap \tilde{N} (p_{2},\epsilon ,k)\right| = 
\sqrt{k+2} (\epsilon -|p_{1}-p_{2}|) \theta (\epsilon -|p_{1}-p_{2}|)+
\mathcal{O}(1) \ .
\end{equation}
By using~\eqref{delta} and choosing
\begin{equation}
\tilde{\alpha} (k)\beta (k) = (k+2)^{1/4} \ ,
\end{equation}
we finally obtain 
\begin{equation}
\left\langle \tilde{\Phi}_{p_{1}}
(z_{1},\bar{z}_{1})\tilde{\Phi}_{p_{2}} (z_{2},\bar{z}_{2})\right\rangle
= \delta (p_{1}-p_{2}) \frac{1}{|z_{1}-z_{2}|^{4h (p_{1})}}  \ .
\end{equation}
A similar analysis can be done for the Ramond fields.

\subsection{Boundary conditions}

The one-point functions of the fields $\tilde{\Phi}_{p}$ are zero for
the discrete class of boundary conditions that we analysed before. For
the continuous class the one-point functions oscillate strongly with
the label $l$, and after averaging they tend to zero as well. The
construction of the chargeless fields however also suggests another
way of defining boundary conditions in the limit theory. Namely we can
scale the boundary labels in analogy to the charge zero fields
$\tilde{\Phi}$,
\begin{equation}\label{AP}
\tilde{B}_{k} (P) = (\lfloor\sqrt{k+2}P\rfloor ,0,0) \ . 
\end{equation}
For such boundary conditions we find trivial one-point functions for
the fields $\Phi_{q,n}$,
\begin{equation}
\langle \Phi_{q,n} (z,\bar{z})\rangle_{P} = 0 \ ,
\end{equation}
but a non-trivial result for the fields $\tilde{\Phi}_{p}$,
\begin{equation}
\langle \tilde{\Phi}_{p} (z,\bar{z})\rangle_{P}
= \frac{\sin (\pi pP)}{\sqrt{\pi p}} |z-\bar{z}|^{-2h} \ .
\end{equation}
Similarly we find in the Ramond sector
\begin{equation}
\langle \Psi^{0}_{q,n}\rangle_{P} = 0 = \langle \Psi^{\pm}_{q,n}\rangle \ ,
\end{equation}
and
\begin{equation}
\left\langle \tilde{\Psi}^{\pm}_{p}\right\rangle_{P} = \frac{\sin (\pi pP)}{\sqrt{\pi
p}} \ .
\end{equation}
These positive results are encouraging to continue the analysis of the
fields $\tilde{\Phi}_{p}$. 

We can also check whether we can define sensible B-type boundary
conditions. They only couple to fields with opposite left- and
right-moving $U (1)$ charges. As we are considering a diagonal theory
with equal left- and right-moving quantum numbers, only fields of
charge zero can couple to a B-type boundary condition in our case. In
the minimal models the B-type boundary
conditions~\cite{Maldacena:2001ky} are labelled only by two labels
$L,S$, where $0\leq L\leq k$ and $S$ is identified modulo $2$. The
boundary states are built from the B-type Ishibashi states by
\begin{equation}
|L,S\rangle = \sum_{l} (2k+4)^{1/2} \frac{S_{(L,0,0)
(l,0,0)}}{\sqrt{S_{(L,0,0) (l,0,0)}}} (-1)^{l/2}
\left( |l,0,0\rangle \!\rangle_{B} + e^{-i\pi S}  |l,0,2\rangle \!\rangle_{B}
\right)  \ .
\end{equation}
In particular this means that we have the one-point functions
\begin{equation}
\langle \phi_{l,m} (z,\bar{z}) \rangle_{(L,S)} = \sqrt{2} \,
\frac{\sin \frac{\pi (l+1) (L+1)}{k+2}}{\sqrt{\sin
\frac{\pi (l+1)}{k+2}}} (-1)^{l/2}\,\delta_{m,0} 
|z-\overline{z}|^{-2 h_{l,0}}\ . 
\end{equation}
Keeping the boundary label fixed while taking the limit, the one-point
function vanishes because of the oscillating sign $(-1)^{l/2}$. On the
other hand, if we redefined the fields $\phi_{l,0}$ by the factor
$(-1)^{l/2}$, we would obtain modified fields $
\tilde{\Phi}_{p}^{(\text{mod})}$ in the limit theory with finite
one-point functions
\begin{equation}\label{Btypeformodfields}
\langle \tilde{\Phi}_{p}^{(\text{mod})}
(z,\bar{z})\rangle_{(L,S)} = \sqrt{2\pi p}\, (L+1) |z-\bar{z}|^{-2h} \ .
\end{equation}
Obviously, we have the relation
\begin{equation}
\langle \ \cdot\ \rangle_{(L,S)} = (L+1) \langle \ \cdot\ \rangle_{(0,S)} \ ,
\end{equation}
which means that all boundary conditions are just superpositions of
the one with $L=0$. Again this reflects the fact that in the minimal
models all B-type boundary conditions can be obtained by a boundary
renormalisation group flow from superpositions of boundary conditions
with $L=0$~\cite{Fredenhagen:2003xf}.

At this point it is hard to decide which of the fields, $\tilde{\Phi}_{p}$ or
$\tilde{\Phi}^{(\text{mod})}_{p}$, is the better definition. 
$\tilde{\Phi}^{(\text{mod})}_{p}$ has a non-trivial one-point function for
B-type boundary conditions, on the other hand its one-point
function in the presence of the A-type
boundary conditions~\eqref{AP} labelled by $P$ vanishes due to
the oscillating sign.

\subsection{Three-point functions}

We now want to determine the three-point correlation functions
involving the fields $\tilde{\Phi}_{p}$, which have charge zero. Due
to charge conservation the correlator $\left\langle \Phi \tilde{\Phi}
\tilde{\Phi}\right\rangle$ is manifestly zero, so we only have to
consider the combinations $\left\langle \Phi \Phi
\tilde{\Phi}\right\rangle$ and $\left\langle
\tilde{\Phi}\tilde{\Phi}\tilde{\Phi}\right\rangle$.

Let us start with the mixed correlator. We have
\begin{align}
&\langle \Phi_{q_{1},n_{1}} (z_{1},\bar{z}_{1})\Phi_{q_{2},n_{2}}
(z_{2},\bar{z}_{2})\tilde{\Phi}_{p} (z_{3},\bar{z}_{3})\rangle \nonumber\\
&\qquad = \lim_{\epsilon \to 0} \lim_{k\to \infty}\beta (k)^{2}\alpha
(k)^{2}\tilde{\alpha} (k) \langle \Phi_{q_{1},n_{1}}^{\epsilon ,k}
(z_{1},\bar{z}_{1}) \Phi_{q_{2},n_{2}}^{\epsilon ,k}
(z_{2},\bar{z}_{2})\tilde{\Phi}_{p}^{\epsilon ,k} (z_{3},\bar{z}_{3})\rangle\\
&\qquad = \lim_{\epsilon \to 0} \lim_{k\to \infty}\frac{\beta (k)^{2}\alpha
(k)^{2}\tilde{\alpha} (k)}{\epsilon^{3} (k+2)^{5/2}} \sum_{\{m_{i}\in
N (q_{i},\epsilon ,k)) \}} \sum_{l\in \tilde{N} (p,\epsilon ,k)}
\delta_{m_{1}+m_{2},0}\nonumber\\
&\qquad \qquad \times C
(|m_{1}|+2n_{1},m_{1};|m_{1}|+2n_{2},-m_{1};l,0) \nonumber\\
&\qquad \qquad \times |z_{12}|^{2(h_{3}-h_{1}-h_{2})}
|z_{13}|^{2(h_{2}-h_{1}-h_{3})}|z_{23}|^{2(h_{1}-h_{2}-h_{3})} \ .
\end{align}
Let us assume that $q_{1}<0$ such that $m_{1}>0$ and $m_{2}=-m_{1}<0$.
To continue we have to determine the asymptotic behaviour of the
three-point coefficient 
\begin{multline}
C(|m_{1}|+2n_{1},m_{1};|m_{1}|+2n_{2},-m_{1};l,0) = 
\begin{pmatrix}
\frac{m_{1}}{2}+n_{1} & \frac{m_{1}}{2}+n_{2} & \frac{l}{2}\\
\frac{m_{1}}{2} & -\frac{m_{1}}{2}& 0
\end{pmatrix}^{\!2}\\
\times 
\sqrt{(|m_{1}|+2n_{1}+1)(|m_{2}|+2n_{2}+1)(l+1)}\, d_{m_{1}+2n_{1},m_{1}+2n_{2},l}\ .
\end{multline}
The coefficient $d_{m_{1}+2n_{1},m_{1}+2n_{2},l}$ behaves as 
\begin{align}
d_{m_{1}+2n_{1},m_{1}+2n_{2},l} & = \frac{\Gamma(1+\rho)}{\Gamma(1-\rho)}P^{2}(m_{1}+n_{1}+n_{2}+\tfrac{l}{2}+1)\nonumber\\
& \quad \times \frac{\Gamma(1-\rho(m_{1}+2n_{1}+1))\Gamma(1-\rho(m_{1}+2n_{2}+1))\Gamma(1-\rho(l+1))}{\Gamma(1+\rho(m_{1}+2n_{1}+1))\Gamma(1+\rho(m_{1}+2n_{2}+1))\Gamma(1+\rho(l+1))}\nonumber\\
&\quad \times
\frac{P^{2}(-n_{1}+n_{2}+\tfrac{l}{2})P^{2}(n_{1}-n_{2}+\tfrac{l}{2})P^{2}(m_{1}+n_{1}+n_{2}-\frac{l}{2})}{P^{2}(m_{1}+2n_{1})P^{2}(m_{1}+2n_{2})P^{2}(l)}
\ .
\end{align}
For the asymptotics of the functions $P$ we use the result of
appendix~\ref{app:Pasymptotics}. From~\eqref{Pasymptotics} we get
\begin{equation}
\frac{P(m_{1}+n_{1}+n_{2}+\tfrac{l}{2}+1)P(m_{1}+n_{1}+n_{2}-\tfrac{l}{2})}{P(m_{1}+2n_{1})P(m_{1}+2n_{2})}
\to  \frac{\Gamma (1+|q_{1}|)}{\Gamma (1-|q_{1}|)}e^{\frac{1}{4}p^{2} 
(\psi(1+|q_{1}|)+\psi (1-|q_{1}|))} \ ,
\end{equation}
where $\psi (x)=\frac{\Gamma ' (x)}{\Gamma (x)}$ is the Digamma
function. Similarly
\begin{equation}
\frac{P(-n_{1}+n_{2}+\tfrac{l}{2})P(n_{1}-n_{2}+\tfrac{l}{2})}{P(l)}
\to \exp \left(\frac{p^{2}\gamma }{2} \right)\ ,
\end{equation}
where $\gamma =\psi (1)$ denotes the Euler-Mascheroni constant. The
coefficient $d_{m_{1}+2n_{1},m_{1}+2n_{2},l}$ thus has the limit
\begin{align}
d_{m_{1}+2n_{1},m_{1}+2n_{2},l} \to e^{\frac{1}{2}p^{2} 
(2\gamma + \psi(1+|q_{1}|)+\psi (1-|q_{1}|))} \ .
\end{align}
The 3j-symbol behaves as (see~\eqref{secondasympof3j} in
appendix~\ref{app:3jsymbols})
\begin{multline}
\begin{pmatrix}
\frac{m_{1}}{2}+n_{1} & \frac{m_{1}}{2}+n_{2} & \frac{l}{2}\\
\frac{m_{1}}{2} & -\frac{m_{1}}{2}& 0
\end{pmatrix}^{\!2} = \frac{\left(n_{1}!n_{2}! \right)^{-1}}{|m_{1}|} 
\left(\frac{4|q_{1}|}{p^{2}} \right)^{\!-(n_{1}+n_{2})}
\left[{}_{2}F_{0} \left(
-n_{1},-n_{2};-\frac{4|q_{1}|}{p^{2}}\right) \right]^{2}\\
\times \left(1+\mathcal{O}
(k^{-1/2}) \right) \ .
\end{multline}
In total the three-point coefficient $C$ has the behaviour
\begin{equation}
C(|m_{1}|+2n_{1},m_{1};|m_{1}|+2n_{2},-m_{1};l,0) \sim
(k+2)^{1/4} \mathcal{C}_{1} (q_{1},n_{1},n_{2},p) \, 
\end{equation}
with the regular function $\mathcal{C}_{1}$ given by
\begin{multline}
\mathcal{C}_{1} (q_{1},n_{1},n_{2},p) =  e^{\frac{1}{2}p^{2} 
(2\gamma + \psi(1+|q_{1}|)+\psi (1-|q_{1}|))}
\left(n_{1}!n_{2}! \right)^{-1} p^{\frac{1}{2}}
\left(\frac{4|q_{1}|}{p^{2}} \right)^{\!-(n_{1}+n_{2})}\\
\times \left[{}_{2}F_{0} \left(
-n_{1},-n_{2};-\frac{4|q_{1}|}{p^{2}}\right) \right]^{2} \ .
\end{multline}
We finally arrive at the following expression for the three point function:
\begin{multline}
\langle \Phi_{q_{1},n_{1}} (z_{1},\bar{z}_{1})\Phi_{q_{2},n_{2}}
(z_{2},\bar{z}_{2})\tilde{\Phi}_{p} (z_{3},\bar{z}_{3})\rangle =\lim_{k\to \infty}\frac{\beta (k)^{2}\alpha
(k)^{2}\tilde{\alpha} (k)}{(k+2)^{3/4}} \\
\times \delta (q_{1}+q_{2})
\mathcal{C}_{1} (q_{1},n_{1},n_{2},p) 
 |z_{12}|^{2(h_{3}-h_{1}-h_{2})}
|z_{13}|^{2(h_{2}-h_{1}-h_{3})}|z_{23}|^{2(h_{1}-h_{2}-h_{3})} \ .
\end{multline}
This three-point function encodes the coupling between two fields
$\Phi$ and one field $\tilde{\Phi}$. It still contains the
normalisation factors. The factor in front can be evaluated as
\begin{equation}
\frac{\beta (k)^{2}\alpha
(k)^{2}\tilde{\alpha} (k)}{(k+2)^{3/4}} = (k+2)^{-1/2}  \ .
\end{equation}
The correlator is therefore suppressed, and fields $\tilde{\Phi}_{p}$ cannot appear in the
operator product expansion of two fields $\Phi_{q_{i},n_{i}}$. Hence, the fields
$\tilde{\Phi}_{p}$ decouple from the fields $\Phi_{q,n}$. Had we
instead used the modified fields $\tilde{\Phi}^{(\text{mod})}_{p}$ that we introduced
before eq.\ \eqref{Btypeformodfields}, we would
have encountered an additional suppression from the oscillating factor
$(-1)^{l/2}$.
\medskip

Next we look at the correlator involving only $\tilde{\Phi}$. We find
\begin{align}
&\left\langle \tilde{\Phi}_{p_{1}} (z_{1},\bar{z}_{1})\tilde{\Phi}_{p_{2}}
(z_{2},\bar{z}_{2})\tilde{\Phi}_{p_{3}} (z_{3},\bar{z}_{3})\right\rangle \nonumber\\
&\qquad = \lim_{\epsilon \to 0} \lim_{k\to \infty}\beta(k)^{2}
\tilde{\alpha} (k)^{3} \left\langle \tilde{\Phi}_{p_{1}}^{\epsilon ,k}
(z_{1},\bar{z}_{1}) \tilde{\Phi}_{p_{2}}^{\epsilon ,k}
(z_{2},\bar{z}_{2})\tilde{\Phi}_{p_{3}}^{\epsilon ,k} (z_{3},\bar{z}_{3})\right\rangle\\
&\qquad = \lim_{\epsilon \to 0} \lim_{k\to \infty}
\frac{\beta(k)^{2}\tilde{\alpha}(k)^{3}}{\epsilon^{3} (k+2)^{5/2}} 
\sum_{\{l_{i}\in \tilde{N} (p_{i},\epsilon ,k)\}} C (\{l_{i},0 \})\nonumber\\
&\qquad \qquad \times |z_{12}|^{2(h_{3}-h_{1}-h_{2})}
|z_{13}|^{2(h_{2}-h_{1}-h_{3})}|z_{23}|^{2(h_{1}-h_{2}-h_{3})} \ .
\end{align}
The three-point coefficient $C$ is given by
\begin{equation}
C (\{l_{i},0 \}) = \begin{pmatrix}
\frac{l_{1}}{2} & \frac{l_{2}}{2} & \frac{l_{3}}{2}\\
0 & 0 & 0
\end{pmatrix}^{\!2} \sqrt{(l_{1}+1) (l_{2}+1) (l_{3}+1)}
d_{l_{1},l_{2},l_{3}}\ .
\end{equation}
Using the asymptotic formula~\eqref{Pasymptotics} for the function $P$ that occurs in the
coefficients $d$, we find
\begin{equation}
d_{l_{1},l_{2},l_{3}} \to 1 \ .
\end{equation} 
The average\footnote{Note that the 3j-symbol in question oscillates
rapidly if one varies the $l_{i}$. Due to the summation over the $l_{i}$
we can insert the average value of the 3j-symbol squared.} of the square of the 3j-symbol behaves as
(see~\eqref{thirdasympof3j})
\begin{multline}
\begin{pmatrix} l_{1}/2& l_{2}/2 & l_{3}/2 \\
0 & 0 & 0
\end{pmatrix}^{\!2}_{\!\text{av}} \sim (k+2)^{-1}\frac{4}{\pi}\\
\times 
 \left((p_{1}+p_{2}+p_{3}) (-p_{1}+p_{2}+p_{3})
(p_{1}-p_{2}+p_{3}) (p_{1}+p_{2}-p_{3}) \right)^{-1/2} \ .
\end{multline}
Notice that the 3j-symbol vanishes if the argument of the square
root becomes negative.
In total the three-point coefficient $C$ becomes
\begin{equation}
C (\{l_{i},0 \}) \sim (k+2)^{-1/4} \mathcal{C}_{2} (\{p_{i} \}) \ ,
\end{equation}
with $\mathcal{C}_{2}$ given by
\begin{equation}
\mathcal{C}_{2}(\{p_{i} \}) = \frac{4}{\pi} \left(\frac{p_{1}p_{2}p_{3}}{(p_{1}+p_{2}+p_{3}) (-p_{1}+p_{2}+p_{3})
(p_{1}-p_{2}+p_{3}) (p_{1}+p_{2}-p_{3})}\right)^{\frac{1}{2}} \ .
\end{equation}
The limit of the three-point function is
\begin{multline}
\left\langle \tilde{\Phi}_{p_{1}} (z_{1},\bar{z}_{1})\tilde{\Phi}_{p_{2}}
(z_{2},\bar{z}_{2})\tilde{\Phi}_{p_{3}} (z_{3},\bar{z}_{3})\right\rangle 
=\lim_{k\to \infty}
\frac{\beta(k)^{2}\tilde{\alpha}(k)^{3}}{ (k+2)^{1/4}} \\
\times \mathcal{C}_{2} (\{p_{i} \})
|z_{12}|^{2(h_{3}-h_{1}-h_{2})}
|z_{13}|^{2(h_{2}-h_{1}-h_{3})}|z_{23}|^{2(h_{1}-h_{2}-h_{3})} \ .
\label{tildephithreepoint}
\end{multline}
Again we encounter a problem with the global factor that is given by
\begin{equation}
\frac{\beta(k)^{2}\tilde{\alpha}(k)^{3}}{ (k+2)^{1/4}} = 
(k+2)^{-1/2} \ ,
\end{equation}
so that also this three-point function is suppressed. Note that we
would have obtained the same result for the modified fields
$\tilde{\Phi}^{(\text{mod})}_{p}$ that we introduced before eq.\
\eqref{Btypeformodfields}, because the 3j-symbol involved is non-zero
only for $l_{1}+l_{2}+l_{3}$ even, so that the sign
$(-1)^{l_{1}+l_{2}+l_{3}}$ that appears in the computation of the
correlator of the three fields $\tilde{\Phi}^{(\text{mod})}_{p}$ is
trivial.

How should we interpret these results? The vanishing of the
three-point function $\langle \tilde{\Phi}\Phi \Phi\rangle$ tells us
that the fields $\tilde{\Phi}$ and $\Phi$ decouple -- in the operator
product expansion (OPE) of two fields $\Phi$ there will never appear a
field $\tilde{\Phi}$, and on the other hand in the OPE of two fields
$\tilde{\Phi}$ there will never be a field $\Phi$ because of charge
conservation. This means that there might be two different limiting
theories, one involving the fields $\Phi$ and one that includes the
fields $\tilde{\Phi}$. If this is true, then we can use a different
normalisation factor $\beta$ for the vacuum in the theory of the
fields $\tilde{\Phi}$, thus rendering the three-point function
$\langle \tilde{\Phi}\tilde{\Phi}\tilde{\Phi}\rangle$ finite. This
will be further explored in~\cite{Fredenhagen:inpreparation}.

\section{Conclusions}

In this article we have analysed the limit of $N=(2,2)$ minimal models
at central charge $c=3$. In the Neveu-Schwarz sector we have
identified fields $\Phi_{q,n}$ that are labelled by their non-zero $U
(1)$ charge $q$ ($0<|q|<1$) and by a discrete label $n\geq 0$. We have
computed the three-point functions of such fields by taking an
appropriate limit of the correlators in the minimal models. We have
also identified boundary conditions in the limit theory that lead to
well-defined disc one-point functions for the fields
$\Phi_{q,n}$. Although we have not checked crossing symmetry of the
three-point functions, our results strongly suggest
that the limit theory exists as a consistent conformal field
theory.

In section~\ref{sec:chargezero} we discussed the question whether
there are additional fields of zero charge. Our results
indicate that there could be such fields $\tilde{\Phi}_{p}$, but they
completely decouple from the charged fields $\Phi_{q,n}$. This points
towards the existence of a second limit theory containing only
chargeless fields. This second theory would arise by a different limit
procedure where in addition to the weight $h$ the label $m$ is kept
fixed. The simplicity of the three-point
function~\eqref{tildephithreepoint} suggests that this second limit
theory might well be the theory of two free bosons and fermions.
\smallskip

It is interesting to compare the results to the less supersymmetric
situations. In a recent article~\cite{Gaberdiel:2011aa},
Gaberdiel and Suchanek argued that the limit of Virasoro minimal
models at central charge $c=1$ (the Runkel-Watts
theory~\cite{Runkel:2001ng}) can be understood as a continuous
orbifold of a free compact boson. A similar construction is proposed
for other limit theories that are based on families of diagonal
cosets. These results suggest that such limit theories could in
general be related to free theories, and that the kind of
non-rationality that one encounters in such limits is similar to the
non-rationality that arises from the existence of a continuum of
twisted sectors. Although the construction of Gaberdiel and Suchanek
cannot be applied directly to the $N=2$ case, because the coset
structure is different, one might still suspect that the limit theory
is related to a free orbifold. We plan to investigate this point in a
subsequent publication~\cite{Fredenhagen:inpreparation}.

In less supersymmetric situations, it has turned out that the
limit theories are related to Liouville or more general conformal Toda
theories. In~\cite{Schomerus:2003vv,Fredenhagen:2004cj} it was shown
that the limit of Virasoro minimal models coincides with the $c=1$
limit of Liouville theory; similarly, the limit of $N=1$ minimal
models is related to $N=1$ Liouville theory~\cite{Fredenhagen:2007tk},
and the limit of $W_{n}$ minimal models to $SU (n)$ conformal Toda
theories~\cite{Fredenhagen:2010zh}. One might therefore wonder whether
the $N=2$ limit theories are related to $N=2$ Liouville theory (see
e.g.\ \cite{Hosomichi:2004ph}) -- or equivalently to its
mirror~\cite{Giveon:1999px,Hori:2001ax}, the supersymmetric
``cigar''. When one compares these theories, one can observe that
the so-called discrete representations in the Liouville spectrum precisely
reproduce the spectrum of our limit theory. It would be interesting
to work out this relation further.
\smallskip

Further clarification of the $N=2$ limit theory will also come from a
geometric point of view. In~\cite{Maldacena:2001ky} a sigma model
interpretation of the minimal models is given, which makes it possible
to understand the limit of large levels also geometrically. This will
be analysed in a forthcoming
publication~\cite{Fredenhagen:inpreparation}. Finally one might
consider the limit also in the framework of Landau-Ginzburg models,
for which a similar limit has been mentioned
in~\cite{Dijkgraaf:1990qw}. It would be interesting to work this out
in more detail.

Limits of $N=2$ models recently have been
discussed~\cite{Creutzig:2011fe,Candu:2012jq} in the context of a
duality of supersymmetric higher spin theories on $AdS_{3}$
backgrounds and two dimensional superconformal theories. There one
does not only take the level $k$ to infinity, but also the label $n$
of the coset $SU(n+1)_{k}/U (n)$ (the minimal models correspond to
$n=1$). Taking first $k\to\infty$ and then $n$ corresponds to the case
of zero 't~Hooft coupling. It would be interesting to extend our
analysis also to the case of $n>1$.

\subsection*{Acknowledgements}

We thank Bianca Dittrich, Matthias Gaberdiel, Ilarion Melnikov, Daniele Oriti and
Volker Schomerus for useful discussions and valuable comments.

\appendix

\section{Asymptotics of 3j-symbols}
\label{app:3jsymbols}
We want to approximate the Wigner 3j-symbols in the limit of large
quantum numbers, in a specific range of parameters defined by the
limiting procedure which is described in the core of this paper.

\subsection{Notations and preliminaries}
To set up our notations, let us briefly state the definition of the
Clebsch-Gordan coefficients. A spin $j$ representation $V_{j}$ of $su (2)$
with standard generators $J_{i}$ satisfying
$[J_{i},J_{j}]=i\epsilon_{ijk}J_{k}$ has a natural basis consisting of
the eigenvectors $|j,\mu\rangle$ of the generator $J_{3}$ with
eigenvalue $\mu$. The tensor product of two irreducible
representations can be decomposed into irreducible representations of
the diagonal subalgebra,
\begin{equation}
V_{j_{1}}\otimes V_{j_{2}} = \bigoplus_{j} V_{j} \ ,
\end{equation}
where $|j_{1}-j_{2}|\leq j\leq j_{1}+j_{2}$ and $j+j_{1}+j_{2}$ is an integer.
The Clebsch-Gordan coefficients 
\begin{equation}
\langle j_1,\mu_1;j_2,\mu_2|j_1,j_2,j,\mu\rangle
\end{equation} 
are then given by the overlap of the two natural sets of basis vectors.

Closely related are the Wigner 3j-symbols that are defined as
\begin{equation}
\begin{pmatrix} j_{1}& j_{2} & j_{3} \\
\mu_{1} & \mu_{2} & \mu_{3}
\end{pmatrix}
:=\frac{(-1)^{j_1-j_2-\mu_3}}{\sqrt{2j_3+1}}\langle
j_1,\mu_1;j_2,\mu_2|j_1,j_2,j_3,-\mu_3\rangle \ ,
\end{equation}
with the choice of conventions: $\mu_3=-\mu =-\mu_1-\mu_2$.\\
An explicit expression was obtained by Racah in~\cite{Racah:1942II}
(see e.g.\ \cite[section 8.2, eq.3]{Varsalovic:book}),
\begin{align}
&\begin{pmatrix} j_{1}& j_{2} & j_{3} \\
\mu_{1} & \mu_{2} & \mu_{3}
\end{pmatrix} =(-1)^{j_{1}-j_{2}-\mu_{3}}
\left(\frac{(j_1\!+\!j_2\!-\!j_{3})!(j_1\!-\!j_2\!+\!j_{3})!(-j_1\!+\!j_2\!+\!j_{3})!}{(j_1\!+\!j_2\!+\!j_{3}\!+\!1)!}\right)^{\!1/2}\nonumber\\
&\qquad \times
[(j_1\!+\!\mu_1)!(j_1\!-\!\mu_1)!(j_2\!+\!\mu_2)!(j_2\!-\!\mu_2)!(j_{3}\!+\!\mu_{3})!
(j_{3}\!-\!\mu_{3})!]^{1/2}  \nonumber \\
&\qquad \times
\sum_z  
\frac{(-1)^z}{z!(j_1\!+\!j_2\!-\!j_{3}\!-\!z)!(j_1\!-\!\mu_1\!-\!z)!
(j_2\!+\!\mu_2\!-\!z)!(j_{3}\!-\!j_2\!+\!\mu_1\!+\!z)!(j_{3}\!-\!j_1\!-\!\mu_2\!+\!z)!
}\;,
\label{Racah}
\end{align}
where the sum over $z$ runs over all the values for which the
arguments of the factorials in the denominator are non-negative. In
particular, this formula provides a simple expression if one of the
labels $\mu_{i}$ is extremal, e.g.\ 
\begin{align}
&\begin{pmatrix}
j_{1} & j_{2} & j_{3}\\
-j_{1} & \mu_{2} & \mu_{3}
\end{pmatrix} = \begin{pmatrix}
j_{3} & j_{1} & j_{2} \\
\mu_{3} & -j_{1} & \mu_{2}
\end{pmatrix}\nonumber\\
& \quad = (-1)^{j_{3}-j_{1}-\mu_{2}}
\left(\frac{(-j_{1}\!+\!j_{2}\!+\!j_{3})!(j_{3}\!+\!\mu_{3})!(j_{2}\!+\!\mu_{2})!(2j_{1})!}{(j_{1}\!-\!j_{2}\!+\!j_{3})!(j_{1}\!+\!j_{2}\!-\!j_{3})!(j_{3}\!-\!\mu
_{3})!(j_{2}\!-\!\mu_{2})!(j_{1}\!+\!j_{2}\!+\!j_{3}\!+\!1)!}
\right)^{\!\frac{1}{2}} \ .
\label{app:extremal3j}
\end{align}

\subsection{Wigner's estimate} 

\begin{figure}[h] \centering \subfloat[][The shaded region is the
projection of the triangle formed by the classical vectors on the
$x$-$y$ plane.]
{\includegraphics[width=.45\columnwidth]{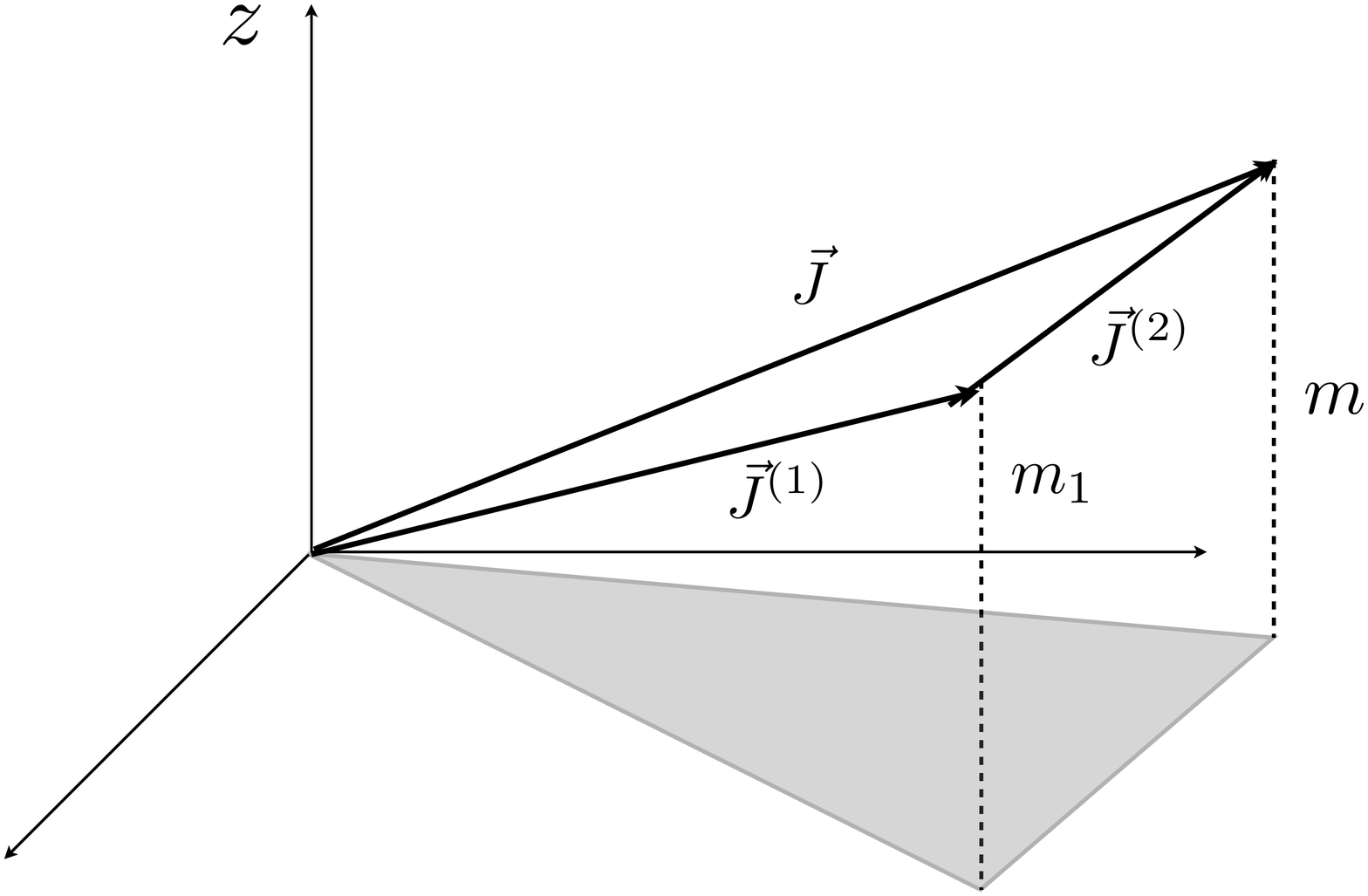}} \quad
\subfloat[][The blue line is a plot coming from the Wigner
estimate~\eqref{Wigner's-estimate}. The points connected by dashed
lines are the exact values of the 3j-symbols. $n$ ranges from 0 to
200.]
{\includegraphics[width=.45\columnwidth]{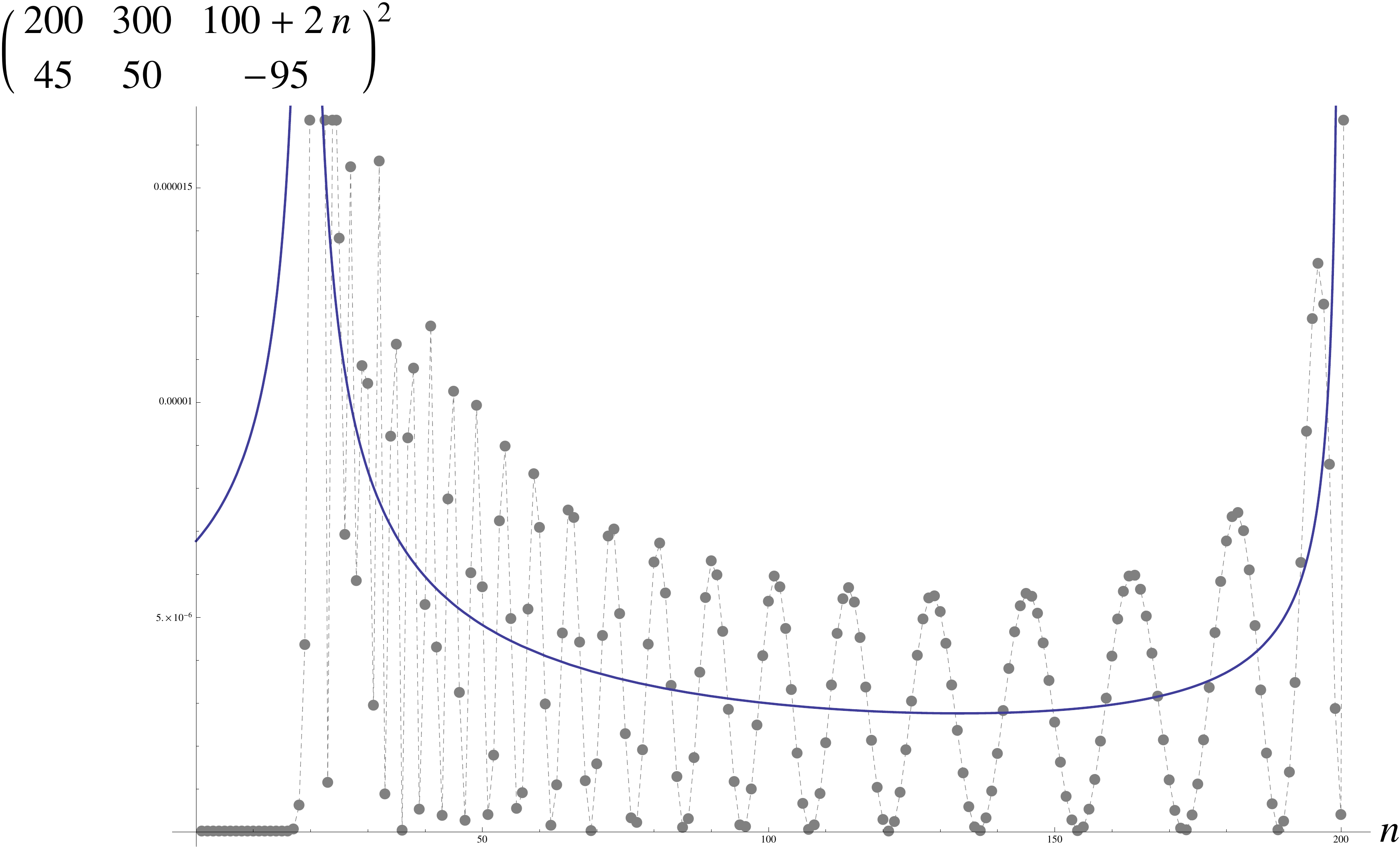}}
\caption{Wigner approximation.}  \label{Wigner-fig}
\end{figure}

For large quantum numbers one expects the Clebsch-Gordan coefficients
to be related to the classical problem of adding angular momenta. This
issue has first been discussed by Wigner in~\cite{Wigner:book}. To
each quantum angular momentum specified by $j_{i},\mu_{i}$ we
therefore associate a vector $\vec{J}^{{(i)}}$ of length squared
$|\vec{J}^{(i)}|^{2}=j (j+1)$ and with specified $z$-component
$J_{z}^{(i)}=\mu_{i}$. The $x$- and $y$- component are not
specified. Classically such angular momenta can be coupled to zero if
they satisfy the condition
$\vec{J}^{(1)}+\vec{J}^{(2)}+\vec{J}^{(3)}=0$. If this is the case,
the triangle their projections form in the $x$-$y$-plane (see
figure~\ref{Wigner-fig} (a)) has an area
\begin{equation}\label{app:defA}
A=\frac{1}{4}\sqrt{(\lambda_1+\lambda_2+\lambda_3) (-\lambda_1+\lambda_2+\lambda_3) (\lambda_1-\lambda_2+\lambda_3) (\lambda_1+\lambda_2-\lambda_3)} \ ,
\end{equation}
where
$\lambda_{i}=\sqrt{|\vec{J}^{(i)}|^{2}-|J_{z}^{(i)}|^{2}}=\sqrt{j_{i}
(j_{i}+1)-\mu_{i}^{2}}$ are the lengths of the projections of
$\vec{J}^{(i)}$ in the $x$-$y$-plane.
The quantum numbers are then said to lie in a \emph{classically
allowed} region. If there are no associated vectors that can be added
to zero, they belong to a \emph{classically forbidden} region; in that
case the ``area'' $A$ in~\eqref{app:defA} is imaginary. If the
projected triangle degenerates ($A=0$), they are said to be in the
\emph{transition region}.

Wigner gave an estimate of the averaged semiclassical behaviour of the
Clebsch-Gordan coefficient in the allowed region~\cite{Wigner:book},
\begin{equation}\label{Wigner's-estimate} |\langle
j_1,\mu_1;j_2,\mu_2|j_1,j_2,j,\mu\rangle|^2_{\text{averaged}}\approx\frac{2j+1}{4\pi|A|}\ .
\end{equation} 
One naturally expects (and it is shown numerically e.g.\ 
in~\cite{Schulten:1971yv}) that the accuracy
of the approximation goes down when the area $A$ is small compared to
the typical length squared of the vectors $\vec{J}^{(i)}$. For more
discussions of the semi-classical asymptotics of the Wigner 3j-symbols
see e.g.\ \cite{Schulten:1971yv,Reinsch:1999}.

For our main application, namely to determine the limit of the
three-point function for the fields $\Phi_{q,n}$, we will see that we
are precisely in this transition region, and we have to follow a
different route to deal with the limit. For the correlator of the
fields $\tilde{\Phi}_{p}$, however, we are in the classically allowed
region, and the Wigner estimate applies. 

\subsection{Asymptotics for the correlators of charged fields} 

When we discuss the limit of the three-point functions for the charged
fields $\Phi_{q_{i},n_{i}}$ we are led to consider the asymptotics of
the 3j-symbols for quantum numbers\footnote{In the main text we use
$l_{i}=2j_{i}$ and $m_{i}=2\mu_{i}$.} $j_{i}=|\mu_{i}|+n_{i}$, where
$n_{i}$ is kept fixed, and the $|\mu_{i}|$ grow linearly in a
parameter $k$.

The 3j-symbol vanishes unless the usual conditions on the addition of
angular momenta are satisfied, namely 
\begin{equation}
\mu_1+\mu_2+\mu_3=0 \quad \text{and}\quad j_{i_{1}}+j_{i_{2}}\geq j_{i_{3}}
\end{equation}
for any permutation $i_{1},i_{2},i_{3}$ of $1,2,3$. If we assume
$\mu_{1},\mu_{2}>0$ and $\mu_{3}<0$, then for large $|\mu_{i}|$ the
conditions on the $j_{i}$ reduce to one condition 
$n_1+n_2\geq n_{3}$.

Because the $z$-components of the angular momenta $\vec{J}^{(i)}$ are
close to maximal in our case, their projections to the $x$-$y$-plane are
short and have lengths 
\begin{equation}
\lambda_{i}=\sqrt{|\mu_{i}| (2n_{i}+1)+n
(n+1)} \ ,
\end{equation}
which only grow with the square root of $k$. This means that
the quantity $A$ given in~\eqref{app:defA}, which describes the area
of the triangle in the $x$-$y$-plane provided it exists, is relatively
small. Thus we are in the transition region between the
classically allowed and the classically forbidden region, and cannot
use the classical Wigner estimate.

Instead we can get the asymptotic behaviour directly from the Racah
formula~\eqref{Racah}. Firstly, we have to understand the
range of $z$ in the sum in~\eqref{Racah}. In the limit of large
$\mu_{i}$ we see that the arguments $j_2+\mu_2-z$ and $j-j_2+\mu_1+z$
do not constrain the sum since they are both surely positive. Bounds
to the summation range are given by the other factorials in the denominator
of equation~\eqref{Racah}, and the summation range is
\begin{equation}
\mathcal{I}:= \{z\in\mathbb{Z}\,|\, z\geq 0,\  z\geq n_1-n_3 ,\ z\leq n_1+n_2-n_3,\
z\leq n_1\}\ .
\end{equation}
Even in the limit of large $|\mu_{i}|$ the summmation range stays
finite, and its lower bound is either zero or $n_1-n_3$ depending on
its sign.

The 3j-symbols can be rewritten as
\begin{equation*}
\begin{split}
& \begin{pmatrix} j_{1}& j_{2} & j_{3} \\
\mu_{1} & \mu_{2} & \mu_{3}
\end{pmatrix} = (-1)^{j_{1}-j_{2}-\mu_{3}}\\
&\times
\underbrace{
\left(\frac{[n_1+n_2-n_3]!}{[2 (|\mu_{1}|+|\mu_{2}|)
+n_1+n_2+n_3+1]!}\right)^{\!1/2}
\times\big((n_{1})!(n_2)!(n_3)![2 (|\mu_{1}|+|\mu_{2}|) +n_3]!\big)^{1/2}}_{\mathbf{N}}\\
&\times
\sum_{z\in\mathcal{I}}\underbrace{
\frac{\left([2|\mu_{1}|+n_1-n_2+n_3]![2|\mu_{1}|+n_1]!\right)^{1/2}}{\left([2|\mu
_{1}|+n_3-n_2+z]!\right)^{1/2}\left([2|\mu_{1}|+n_3-n_2+z]!\right)^{1/2}}
}_{\mathbf A}\\
&\qquad \times \underbrace{
\frac{\left([2|\mu_{2}|+n_2-n_1+n_3]![2|\mu_{2}|+n_2]!\right)^{1/2}}{\left([2|\mu
_{2}|+n_2-z]!\right)^{1/2}\left([2|\mu_{2}|+n_2-z]!\right)^{1/2}}
}_{\mathbf B}
\\
& \qquad \times \underbrace{\frac{(-1)^z}{z!}\frac{1}{[n_1+n_2-n_3-z]![n_1-z]![n_3-n_1+z]!}}_{\mathbf C}
\ .\end{split}
\end{equation*}
Using the fact that $k$ is large we are able to recast parts $A$ and $B$ using that
\begin{equation}\label{limitoffactorials}
\frac{(K+a)!}{K!}=\frac{K!}{K!}\times
(K+1)\dots(K+a)=K^a\left(1+\mathcal{O}\left(\frac{1}{K}\right)
\right)\qquad \text{for large}\ K
\end{equation}
so that the leading contributions read
\begin{equation}
A\approx (2|\mu_{1}|)^{n_1+\frac{n_{2}-n_{3}}{2}-z}\ ,\qquad B\approx
(2|\mu_{2}|)^{z-\frac{n_{1}-n_{3}}{2}}\ .
\end{equation}
Similarly, the factor $N$ can be approximated by
\begin{equation}
N\approx (2|\mu_{1}|+2|\mu_{2}|)^{-\frac{n_{1}+n_{2}+1}{2}}\times\sqrt{n_{1}!n_{2}!n_{3}![n_{1}+n_{2}-n_{3}]!}\ .
\end{equation}
The 3j-symbol then reads
\begin{align}
& \begin{pmatrix} j_{1}& j_{2} & j_{3} \\
\mu_{1} & \mu_{2} & \mu_{3}
\end{pmatrix} = (-1)^{2|\mu_{1}|+n_{1}-n_{2}}\sqrt{n_{1}!n_{2}!n_{3}![n_{1}+n_{2}-n_{3}]!}
(2|\mu_{1}|+2|\mu_{2}|)^{-\frac{n_{1}+n_{2}+1}{2}}\nonumber \\
&\quad \times \sum_{z\in\mathcal{I}} (-1)^{z} \frac{1}{z![n_1+n_2-n_3-z]![n_1-z]![n_3-n_1+z]!} 
(2|\mu_{1}|)^{n_1+\frac{n_{2}-n_{3}}{2}-z}(2|\mu_{2}|)^{z-\frac{n_{1}-n_{3}}{2}}\ .
\end{align}
Introducing the notation
\begin{equation}
J= \frac{n_{1}+n_{2}}{2}\quad ,\quad M=\frac{n_{1}-n_{2}}{2}\quad ,\quad 
M' = -\frac{n_{1}+n_{2}}{2}+n_{3}\ ,
\end{equation}
and
\begin{equation}
\cos \beta = \frac{|\mu_{1}|-|\mu_{2}|}{|\mu_{1}|+|\mu_{2}|}\ ,
\end{equation}
we can express the asymptotic form of the 3j-symbol as
\begin{equation}\label{app:3jasymptotic}
\begin{pmatrix} j_{1}& j_{2} & j_{3} \\
\mu_{1} & \mu_{2} & \mu_{3}
\end{pmatrix} \approx (-1)^{2|\mu_{1}|+n_{3}-n_{2}}
(2|\mu_{1}|+2|\mu_{2}|)^{-\frac{1}{2}} d^{J}_{M',M} (\beta) \ .
\end{equation}
Here, $d^{J}_{M',M} (\beta)$ denotes the Wigner d-matrix~\cite{Wigner:book,Varsalovic:book},
\begin{align}
&d^{J}_{M',M} (\beta) = \sqrt{(J\!+\!M')!(J\!-\!M')!(J\!+\!M)!(J\!-\!M)!} \nonumber \\
&\quad \times \sum_{z}
\frac{(-1)^{M'-M+z}}{(J\!+\!M\!-\!z)!z!(M'\!-\!M\!+\!z)!(J\!-\!M'\!-\!z)!}
\left(\cos \tfrac{\beta}{2} \right)^{2J+M-M'-2z}
\left(\sin \tfrac{\beta}{2} \right)^{M'-M+2z} \ .
\end{align}
The Wigner d-matrix is expressible in terms of standard $_{2}F_{1}$
hypergeometric functions. More precisely, for $n_{1}\leq n_{3}$, we find
\begin{align}
&\begin{pmatrix}j_{1}&j_{2}&j_{3}\\
\mu_{1}&\mu_{2}&\mu_{3}\end{pmatrix} = 
(-1)^{2\mu_1+n_{1}-n_{2}} \frac{1}{(n_{3}-n_{1})!} \sqrt{\frac{n_{2}!n_{3}!}{n_{1}!(n_{1}+n_{2}-n_{3})!}}\,(2|\mu_{1}|+2|\mu_{2}|)^{-\frac{1}{2}}\nonumber\\  
& \qquad \times  \frac{|\mu_{1}|^{n_1+\frac{n_2-n_3}{2}}|\mu_{2}|^{\frac{n_{3}-n_{1}}{2}}}{(|\mu_{1}|+|\mu_{2}|)^{\frac{n_{1}+n_{2}}{2}}} \, {}_2F_1\left(n_3-n_2-n_1,-n_1;n_3-n_1+1;-\frac{|\mu_{2}|}{|\mu_{1}|}\right)\left(1+\mathcal{O}(1/k)\right)\ ,
\end{align}
whereas for $n_{1}\geq n_{3}$ we have
\begin{align}
&\begin{pmatrix}j_{1}&j_{2}&j_{3}\\
\mu_{1}&\mu_{2}&\mu_{3}\end{pmatrix} = 
(-1)^{2\mu_1+n_{1}-n_{2}}\frac{1}{(n_{1}-n_{3})!} \sqrt{\frac{n_{1}!}{n_{2}!n_{3}!}}
\,(2|\mu_{1}|+2|\mu_{2}|)^{-\frac{1}{2}}\nonumber\\  
&\qquad \times \frac{|\mu_{1}|^{\frac{n_{2}+n_{3}}{2}}|\mu_{2}|^{\frac{n_{1}-n_{3}}{2}}}{(|\mu_{1}|+|\mu_{2}|)^{\frac{n_{1}+n_{2}}{2}}}\, {}_2F_1\left(-n_3,-n_2;n_{1}-n_{3}+1;-\frac{|\mu_{2}|}{|\mu_{1}|}\right)\left(1+\mathcal{O}(1/k)\right)\ .
\end{align}

\subsection{Asymptotics for the mixed correlators}

Now we want to study the asymptotics of the 3j-symbol when two of the
$(j,\mu)$ pairs behave as before, i.e.\ $j_{i}=|\mu_{i}|+n_{i}$ ($i=1,2$) with
fixed non-negative integers $n_{i}$, and the quantum numbers $\mu_{i}$
grow linearly with the parameter $k$. For the third coloumn we choose
$\mu_{3}=0$ and $j_{3}$ grows with the square root of $k$. As the
labels $\mu_{i}$ have to add up to zero, we have $\mu_{2}=-\mu_{1}$
and we choose $\mu_{1}$ to be positive.

From the Racah formula~\eqref{Racah} we find
\begin{align}
&\begin{pmatrix} j_{1}& j_{2} & j_{3} \\
\mu_{1} & \mu_{2} & 0
\end{pmatrix} = (-1)^{n_{1}-n_{2}} \left(n_{1}!n_{2}! \right)^{1/2}\nonumber\\
& \quad \times \sum_{z} (-1)^{z} \left(
\frac{(2\mu_{1}+n_{1}+n_{2}-j_{3})!(2\mu_{1}+n_{1})!(2\mu_{1}+n_{2})!}{(2\mu
_{1}+n_{1}+n_{2}+j_{3}+1)![(2\mu_{1}+n_{1}+n_{2}-j_{3}-z)!]^{2}}
\right)^{\!1/2}\nonumber\\
& \qquad \times
\left(\frac{(j_{3}+n_{1}-n_{2})!(j_{3}-n_{1}+n_{2})![j_{3}!]^{2}}{[(j_{3}-n_{1}+z)!(j_{3}-n_{2}+z)!]^{2}}
\right)^{\!1/2} \frac{1}{z!(n_{1}-z)!(n_{2}-z)!} \ ,
\end{align}
where the sum runs from $z=0$ to $z=\text{min} (n_{1},n_{2})$. As in
the previous discussion, the ratios of the factorials growing with
$2\mu_{1}$ and also the ratios of the factorials growing with~$j_{3}$
can be approximated by~\eqref{limitoffactorials}, and we obtain
\begin{align}
&\begin{pmatrix} j_{1}& j_{2} & j_{3} \\
\mu_{1} & \mu_{2} & 0
\end{pmatrix}\nonumber\\
& = (-1)^{n_{1}-n_{2}} \frac{\left(n_{1}!n_{2}! \right)^{1/2}}{( 2\mu_{1})^{1/2}} 
\sum_{z} \frac{(-1)^{z}}{z!(n_{1}-z)!(n_{2}-z)!}
\left(\frac{2\mu_{1}}{j_{3}^{2}} \right)^{\!z-\frac{n_{1}+n_{2}}{2}}\left(1+\mathcal{O} (k^{-1/2}) \right) \\
& = (-1)^{n_{1}-n_{2}} \frac{\left(n_{1}!n_{2}! \right)^{-1/2}}{( 2\mu_{1})^{1/2}} 
\left(\frac{2\mu_{1}}{j_{3}^{2}} \right)^{\!-\frac{n_{1}+n_{2}}{2}}
{}_{2}F_{0} \left( -n_{1},-n_{2};-\frac{2\mu_{1}}{j_{3}^{2}}\right)
\left(1+\mathcal{O} (k^{-1/2}) \right)\ ,
\label{secondasympof3j}
\end{align}
where ${}_{2}F_{0}$ denotes the corresponding hypergeometric function.

\subsection{Asymptotics for the correlators of uncharged fields}

Our third region of interest has all $\mu_{i}=0$, and the $j_{i}$ are
growing at the same rate, proportional to the square root of $k$. The
corresponding 3j-symbols are given by~\cite[section 8.5, eq.32]{Varsalovic:book}
\begin{align}
\begin{pmatrix} j_{1}& j_{2} & j_{3} \\
0 & 0 & 0
\end{pmatrix} &= (-1)^{\frac{j_{1}+j_{2}+j_{3}}{2}}
\left(\frac{(-j_{1}+j_{2}+j_{3})!(j_{1}-j_{2}+j_{3})!(j_{1}+j_{2}-j_{3})!}{(j_{1}+j_{2}+j_{3}+1)!}\right)^{\!1/2} \nonumber\\
&\quad \times 
\frac{\left(\frac{j_{1}+j_{2}+j_{3}}{2}
\right)!}{\left(\frac{-j_{1}+j_{2}+j_{3}}{2}
\right)!\left(\frac{j_{1}-j_{2}+j_{3}}{2}
\right)!\left(\frac{j_{1}+j_{2}-j_{3}}{2} \right)!} 
\end{align}
if $|j_{1}-j_{2}|\leq j_{3}\leq j_{1}+j_{2}$ and if
$j_{1}+j_{2}+j_{3}$ is an even integer, otherwise it vanishes. 

Let us for a moment assume that $j_{1}+j_{2}+j_{3}$ is even. To
analyse the behaviour of the 3j-symbol we use Stirling's formula for
the factorial,
\begin{equation}
n! = \sqrt{2\pi n}\,n^{n}e^{-n} (1+\mathcal{O} (1/n)) \ .
\end{equation}
We find
\begin{align}
&\begin{pmatrix} j_{1}& j_{2} & j_{3} \\
0 & 0 & 0
\end{pmatrix} =  (-1)^{\frac{j_{1}+j_{2}+j_{3}}{2}}
\sqrt{\frac{2}{\pi}} \nonumber\\
&\quad \times \left((j_{1}+j_{2}+j_{3}) (-j_{1}+j_{2}+j_{3})
(j_{1}-j_{2}+j_{3}) (j_{1}+j_{2}-j_{3}) \right)^{-1/4} (1+\mathcal{O}
(k^{-1/2}))\ .
\end{align}
For the computations in the main text we are interested in the
averaged value of the square of the 3j-symbol. In the allowed region,
i.e.\ where $|j_{1}-j_{2}|\leq j_{3}\leq j_{1}+j_{2}$, every second
3j-symbol vanishes due to the constraint that $j_{1}+j_{2}+j_{3}$
should be even. Therefore we obtain
\begin{equation}
\begin{pmatrix} j_{1}& j_{2} & j_{3} \\
0 & 0 & 0
\end{pmatrix}^{2}_{\text{av}} \approx \frac{1}{\pi}
 \left((j_{1}+j_{2}+j_{3}) (-j_{1}+j_{2}+j_{3})
(j_{1}-j_{2}+j_{3}) (j_{1}+j_{2}-j_{3}) \right)^{-1/2} \ .
\label{thirdasympof3j}
\end{equation}
This precisely equals the Wigner estimate~\eqref{Wigner's-estimate},
with the area $A$ given in~\eqref{app:defA}. 

\section{Asymptotics of products of Gamma functions}
\label{app:Pasymptotics}

The three-point coefficient contains products of Gamma functions of
the form (see~\eqref{def_P})
\begin{equation}
P(l)=\prod_{j=1}^{l}\frac{\Gamma(1+j\rho)}{\Gamma(1-j\rho)} \ ,
\end{equation}
where $\rho =1/ (k+2)$. When we take the limit $k\to \infty$, also the
quantum numbers become large, so that we have to determine the
asymptotics of $P (l)$ for large $l$ and $k$.

We write $l=f/\rho$, where $f$ tends towards a constant $f_{0}$ in the
limit,
\begin{equation}
\lim_{k\to \infty} f = f_{0} \quad ,\quad 0\leq f_{0}<1 \ .
\end{equation}
We then have
\begin{align}
P (f/\rho) &= \exp \left(\sum_{j=1}^{f\rho^{-1}}\log \frac{\Gamma
(1+j\rho)}{\Gamma (1-j\rho)} \right)\\
&= \exp \left(\rho^{-1}\int_{0}^{f}\log \frac{\Gamma (1+x)}{\Gamma
(1-x)}dx +\frac{1}{2}\log \frac{\Gamma (1+f)}{\Gamma (1-f)} +
\mathcal{O} (\rho) \right) \ ,
\end{align}
where we employed the Euler-MacLaurin sum formula (see e.g.\
\cite{Andrews:book}). The integral is given by (see e.g.\ \cite{Barnes:1900})
\begin{equation}
\int_{0}^{f}\log \frac{\Gamma (1+x)}{\Gamma
(1-x)}dx = -f^{2}+f\log \frac{\Gamma (1+f)}{\Gamma (1-f)} -\log
\left[G (1+f)G (1-f) \right] \ ,
\end{equation}
where $G$ is the Barnes G-function\footnote{$G$ is related to the Barnes
double gamma function $\Gamma_{2} (z;b_{1},b_{2})$ by $G
(z)=\sqrt{2\pi} (\Gamma_{2} (z;1,1))^{-1}$.}. 

When we write $f=f_{0}+f_{1}$, where $f_{1}$ goes to zero in the
limit, we obtain the asymptotic formula
\begin{align}
& P (f/\rho ) = \exp \bigg(\rho^{-1}\left(-f_{0}^{2}+f_{0}\log
\frac{\Gamma (1+f_{0})}{\Gamma (1-f_{0})}-\log \left[G (1+f_{0})G
(1-f_{0}) \right] \right) \nonumber\\
&\qquad  + (\rho^{-1}f_{1} +\tfrac{1}{2})\log \frac{\Gamma
(1+f_{0})}{\Gamma (1-f_{0})} + \frac{\rho^{-1}f_{1}^{2}}{2} \left(\psi
(1+f_{0})+\psi (1-f_{0})\right) +\mathcal{O} (f_{1},\rho
,\rho^{-1}f_{1}^{3}) \bigg) \ .
\label{Pasymptotics}
\end{align}
Here, $\psi (x)=\frac{\Gamma ' (x)}{\Gamma (x)}$ denotes the Digamma function.

\section{Odd channel three-point functions}
\label{app:oddchannelthreepointfunctions}
In this section we consider three-point functions of two primaries and
one superdescendant field in minimal models. In the coset model
description they can be derived from the three-point function of the
$SU (2)$ WZW models as it has been done for the three-point function
of three primaries in~\cite{Mussardo:1988av}. To this end one has to realise the
superdescendants explicitly as descendants in the $SU (2)$ model and
determine the corresponding correlators. Although the computation is
straightforward, to our knowledge these results have not appeared in
the literature before.

For explicitness let us consider the Neveu-Schwarz correlator
\begin{equation}\label{app:odd3pt}
\langle (\bar{G}^{+}_{-\frac{1}{2}}G^{+}_{-\frac{1}{2}}\phi_{l_{1},m_{1}}) (z_{1},\bar{z}_{1})
\phi_{l_{2},m_{2}} (z_{2},\bar{z}_{2})\phi_{l_{3},m_{3}}
(z_{3},\bar{z}_{3}) \rangle \ ,
\end{equation}
where we assume that $|m_{i}|\leq l_{i}$ and $m_{1}>0$. Due to charge
conservation a non-zero correlator has to satisfy
\begin{equation}\label{app:charge}
1-\frac{m_{1}}{k+2}-\frac{m_{2}}{k+2}-\frac{m_{3}}{k+2} = 0\ .
\end{equation}
In the coset description we have
\begin{equation}
\bar{G}^{+}_{-\frac{1}{2}}G^{+}_{-\frac{1}{2}}|l,m,0\rangle = \left(\frac{l (l+2)-m (m-2)}{2
(k+2)} \right)^{\!-1} |l,m,2\rangle  
\end{equation}
for $-l+2\leq m\leq l$. Notice that as in the main text we have chosen
the diagonal minimal models with equal holomorphic and
anti-holomorphic quantum numbers ($\bar{m}=m$). To relate the above
three-point function to a correlator in the $SU (2)$ model we have to
use the field identification 
\begin{equation}
|l_{1},m_{1},2\rangle =
|k-l_{1},m_{1}-k-2,0\rangle = |\tilde{l}_{1},-\tilde{l}_{1}-2n_{1}-2,0\rangle \ ,
\end{equation}
where we set $\tilde{l}_{1}=k-l_{1}$ and
$n_{1}=\frac{l_{1}-|m_{1}|}{2}$. Then the $U (1)$ part of the coset
trivially factorises (all three labels $s_{i}$ are $0$, and the
new labels $m_{i}$ add up to zero, $(m_{1}-k-2)+m_{2}+m_{3}=0$, which
corresponds to the charge conservation
condition~\eqref{app:charge}). The coset state
$|\tilde{l}_{1},-\tilde{l}_{1}-2 (n_{1}+1),0\rangle $ comes from the
state
\begin{equation}
\zeta_{l_{1},n_{1}} = \gamma_{l_{1},n_{1}}^{-1} (J^{-}_{-1})^{n_{1}+1} (\bar{J}^{-}_{-1})^{n_{1}+1} |\tilde{l}_{1},-\tilde{l}_{1},-\tilde{l}_{1}\rangle_{SU (2)} 
\end{equation}
in the $SU (2)$ model, where $\gamma_{l_{1},n_{1}}$ is a normalisation
factor to ensure that $|\zeta_{l_{1},n_{1}}|^{2}=1$. The conventions
for the $SU (2)$ current algebra that we use here are given by
\begin{equation}
[J^{+}_{m},J^{-}_{n}] = 2J^{0}_{m+n} + k m \delta_{m+n,0} \quad ,\quad
[J^{0}_{m},J^{\pm}_{n}] = \pm J^{\pm}_{m+n} \ .
\end{equation}
The primary states in the $SU (2)$ model are labelled by
$|l,m,\bar{m}\rangle_{SU (2)}$, where in our conventions
\begin{align}
J^{0}_{0}|l,m,\bar{m}\rangle_{SU (2)} & = \frac{m}{2} |l,m,\bar{m}\rangle_{SU (2)}\\
\left((J^{0}_{0})^{2} +\frac{1}{2} J^{+}_{0}J^{-}_{0} + \frac{1}{2} J^{-}_{0}J^{+}_{0}
\right)|l,m,\bar{m}\rangle_{SU (2)} &= \frac{l (l+2)}{4}
|l,m,\bar{m}\rangle_{SU (2)}\\
J_{0}^{\pm}|l,m,\bar{m}\rangle_{SU (2)} &= c_{\pm} (l,m) |l,m\pm
2,\bar{m}\rangle_{SU (2)} \ ,
\end{align}
where
\begin{equation}
c_{\pm} (l,m)=\frac{1}{2}\sqrt{(l\mp m) (l\pm m+2)} \ .
\end{equation}
Anologous relations hold for the operators $\bar{J}$.
Now it is easy to check inductively that
\begin{equation}
\gamma_{l_{1},n_{1}} = \frac{(n_{1}+1)!\,l_{1}!}{(l_{1}-n_{1}-1)!} \ .
\end{equation}
The coefficient of the three-point function~\eqref{app:odd3pt}
therefore can be read off from the $SU (2)$ correlator
\begin{multline}\label{app:def_F}
F = \left(\frac{2 (n_{1}+1) (l_{1}-n_{1})\gamma_{l_{1},n_{1}}}{(k+2)}
\right)^{\!-1} \\
\times \langle \left((J^{-}_{-1})^{n_{1}+1} (\bar{J}^{-}_{-1})^{n_{1}+1}
\chi_{\tilde{l}_{1},-\tilde{l}_{1},-\tilde{l}_{1}}\right)
(z_{1},\bar{z}_{1}) \chi_{l_{2},m_{2},m_{2}} (z_{2},\bar{z}_{2})
\chi_{l_{3},m_{3},m_{3}} (z_{3},\bar{z}_{3})\rangle \ ,
\end{multline}
where we denoted the field corresponding to the state
$|l,m,\bar{m}\rangle$ by $\chi_{l,m,\bar{m}}$. 
This correlator can be computed starting from the known three-point
function for primary fields~\cite{Zamolodchikov:1986bd,Dotsenko:1990zb},
\begin{multline}
\langle \chi_{l_{1},m_{1},\bar{m}_{1}}
(z_{1},\bar{z}_{1})\chi_{l_{2},m_{2},\bar{m}_{2}}
(z_{2},\bar{z}_{2})\chi_{l_{3},m_{3},\bar{m}_{3}} (z_{3},\bar{z}_{3})\rangle
= 
\begin{pmatrix}
\frac{l_{1}}{2} & \frac{l_{2}}{2} & \frac{l_{3}}{2}\\[1mm]
 \frac{m_{1}}{2} & \frac{m_{2}}{2} & \frac{m_{3}}{2}
\end{pmatrix}
\begin{pmatrix}
\frac{l_{1}}{2} & \frac{l_{2}}{2} & \frac{l_{3}}{2}\\[1mm]
 \frac{\bar{m}_{1}}{2} & \frac{\bar{m}_{2}}{2} & \frac{\bar{m}_{3}}{2}
\end{pmatrix}\\
\times \sqrt{(l_{1}+1) (l_{2}+1) (l_{3}+1)}\,d_{l_{1},l_{2},l_{3}}\,
|z_{12}|^{2 (h_{l_{3}}-h_{l_{1}}-h_{l_{2}})}|z_{13}|^{2
(h_{l_{2}}-h_{l_{1}}-h_{l_{3}})}|z_{23}|^{2 (h_{l_{1}}-h_{l_{2}}-h_{l_{3}})} \ ,
\end{multline}
where $d_{l_{1},l_{2},l_{3}}$ is given in~\eqref{defofd}, and the
conformal weights are
\begin{equation}
h_{l} = \frac{l (l+2)}{4 (k+2)}\ .
\end{equation}
Correlators of descendant fields are then computed by the usual
contour integral techniques. Let us start with the simple case that
there is only one operator $J^{-}_{-1}$ acting on
$\chi_{\tilde{l}_{1},-\tilde{l}_{1},-\tilde{l}_{1}}$. We find
\begin{align}
&\langle \left(J^{-}_{-1} 
\chi_{\tilde{l}_{1},-\tilde{l}_{1},-\tilde{l}_{1}}\right)
(z_{1},\bar{z}_{1}) \chi_{l_{2},m_{2},\bar{m}_{2}} (z_{2},\bar{z}_{2})
\chi_{l_{3},m_{3},\bar{m}_{3}} (z_{3},\bar{z}_{3})\rangle\nonumber\\
&\quad = \frac{1}{2\pi i} \oint_{z_{1}}\frac{dw}{w-z_{1}} \langle J^{-} (w) 
\chi_{\tilde{l}_{1},-\tilde{l}_{1},-\tilde{l}_{1}}
(z_{1},\bar{z}_{1}) \chi_{l_{2},m_{2},\bar{m}_{2}} (z_{2},\bar{z}_{2})
\chi_{l_{3},m_{3},\bar{m}_{3}} (z_{3},\bar{z}_{3})\rangle\nonumber\\
&\quad = - \frac{1}{2\pi i}\left(\oint_{z_{2}} +\oint_{z_{3}}
\right)\frac{dw}{w-z_{1}} \langle J^{-} (w) 
\chi_{\tilde{l}_{1},-\tilde{l}_{1},-\tilde{l}_{1}}
(z_{1},\bar{z}_{1}) \chi_{l_{2},m_{2},\bar{m}_{2}} (z_{2},\bar{z}_{2})
\chi_{l_{3},m_{3},\bar{m}_{3}} (z_{3},\bar{z}_{3})\rangle\nonumber\\
&\quad = \frac{1}{z_{12}} \langle 
\chi_{\tilde{l}_{1},-\tilde{l}_{1},-\tilde{l}_{1}}
(z_{1},\bar{z}_{1}) (J^{-}_{0}\chi_{l_{2},m_{2},\bar{m}_{2}}) (z_{2},\bar{z}_{2})
\chi_{l_{3},m_{3},\bar{m}_{3}} (z_{3},\bar{z}_{3})\rangle\nonumber\\
&\qquad  +\frac{1}{z_{13}} \langle 
\chi_{\tilde{l}_{1},-\tilde{l}_{1},-\tilde{l}_{1}}
(z_{1},\bar{z}_{1}) \chi_{l_{2},m_{2},\bar{m}_{2}} (z_{2},\bar{z}_{2})
(J^{-}_{0}\chi_{l_{3},m_{3},\bar{m}_{3}}) (z_{3},\bar{z}_{3})\rangle\nonumber\\
&\quad = \frac{1}{z_{12}} c_{-} (l_{2},m_{2})  \langle 
\chi_{\tilde{l}_{1},-\tilde{l}_{1},-\tilde{l}_{1}}
(z_{1},\bar{z}_{1}) \chi_{l_{2},m_{2}-2,\bar{m}_{2}} (z_{2},\bar{z}_{2})
\chi_{l_{3},m_{3},\bar{m}_{3}} (z_{3},\bar{z}_{3})\rangle\nonumber\\
&\qquad + \frac{1}{z_{13}} c_{-} (l_{3},m_{3})  \langle 
\chi_{\tilde{l}_{1},-\tilde{l}_{1},-\tilde{l}_{1}}
(z_{1},\bar{z}_{1}) \chi_{l_{2},m_{2},\bar{m}_{2}} (z_{2},\bar{z}_{2})
\chi_{l_{3},m_{3}-2,\bar{m}_{3}} (z_{3},\bar{z}_{3})\rangle \ .
\end{align}
Due to the shift relations of 3j-symbols~\cite[section 8.4,
eq.5]{Varsalovic:book} we have
\begin{equation}
c_{-} (l_{3},m_{3}) \begin{pmatrix}
\frac{\tilde{l}_{1}}{2} & \frac{l_{2}}{2} & \frac{l_{3}}{2} \\[1mm]
-\frac{\tilde{l}_{1}}{2} & \frac{m_{2}}{2} & \frac{m_{3}}{2}-1
\end{pmatrix} = -c_{-} (l_{2},m_{2})  \begin{pmatrix}
\frac{\tilde{l}_{1}}{2} & \frac{l_{2}}{2} & \frac{l_{3}}{2} \\[1mm]
-\frac{\tilde{l}_{1}}{2} & \frac{m_{2}}{2}-1 & \frac{m_{3}}{2}
\end{pmatrix} \ .
\end{equation}
Therefore
\begin{align}
&\langle \left(J^{-}_{-1} 
\chi_{\tilde{l}_{1},-\tilde{l}_{1},-\tilde{l}_{1}}\right)
(z_{1},\bar{z}_{1}) \chi_{l_{2},m_{2},\bar{m}_{2}} (z_{2},\bar{z}_{2})
\chi_{l_{3},m_{3},\bar{m}_{3}} (z_{3},\bar{z}_{3})\rangle\nonumber\\
&\quad = \left(\frac{1}{z_{12}}-\frac{1}{z_{13}} \right) c_{-} (l_{2},m_{2}) 
 \langle 
\chi_{\tilde{l}_{1},-\tilde{l}_{1},-\tilde{l}_{1}}
(z_{1},\bar{z}_{1}) \chi_{l_{2},m_{2}-2,\bar{m}_{2}} (z_{2},\bar{z}_{2})
\chi_{l_{3},m_{3},\bar{m}_{3}} (z_{3},\bar{z}_{3})\rangle\\
&\quad = c_{-} (l_{2},m_{2}) \begin{pmatrix}
\frac{\tilde{l}_{1}}{2} & \frac{l_{2}}{2} & \frac{l_{3}}{2} \\[1mm]
-\frac{\tilde{l}_{1}}{2} & \frac{m_{2}}{2}-1 & \frac{m_{3}}{2}
\end{pmatrix}
\begin{pmatrix}
\frac{\tilde{l}_{1}}{2} & \frac{l_{2}}{2} & \frac{l_{3}}{2}\\[1mm]
 -\frac{\tilde{l}_{1}}{2} & \frac{\bar{m}_{2}}{2} & \frac{\bar{m}_{3}}{2}
\end{pmatrix} \sqrt{\big(\tilde{l}_{1}+1\big) \big(l_{2}+1\big)
\big(l_{3}+1\big)}\,
d_{\tilde{l}_{1},l_{2},l_{3}}\nonumber \\
&\qquad \times 
z_{12}^{h_{l_{3}}- (h_{\tilde{l}_{1}}+1)-h_{l_{2}}}
\bar{z}_{12}^{h_{l_{3}}-h_{\tilde{l}_{1}}-h_{l_{2}}} 
z_{23}^{(h_{\tilde{l}_{1}}+1)-h_{l_{2}}-h_{l_{3}}}
\bar{z}_{23}^{h_{\tilde{l}_{1}}-h_{l_{2}}-h_{l_{3}}}
z_{13}^{h_{l_{2}}- (h_{\tilde{l}_{1}}+1)-h_{l_{3}}}
\bar{z}_{13}^{h_{l_{2}}-h_{\tilde{l}_{1}}-h_{l_{3}}} \ .
\end{align}
Following this procedure iteratively, one obtains an expression for
the correlator given in~\eqref{app:def_F},
\begin{align}
F = &\left(\frac{2 (n_{1}+1) (l_{1}-n_{1})\gamma_{l_{1},n_{1}}}{(k+2)}
\right)^{\!-1} \left(\prod_{i=0}^{n_{1}} c_{-} (l_{2},m_{2}-2i)
\right)^{\!2} \nonumber \\
& \times \begin{pmatrix}
\frac{\tilde{l}_{1}}{2} & \frac{l_{2}}{2} & \frac{l_{3}}{2} \\[1mm]
-\frac{\tilde{l}_{1}}{2} & \frac{m_{2}}{2}-n_{1}-1 & \frac{m_{3}}{2}
\end{pmatrix}^{\!2}
\sqrt{\big(\tilde{l}_{1}+1\big) \big(l_{2}+1\big) \big(l_{3}+1\big)}\, 
d_{\tilde{l}_{1},l_{2},l_{3}}\nonumber\\
& \times 
|z_{12}|^{2 (h_{l_{3}}- (h_{\tilde{l}_{1}}+n_{1}+1)-h_{l_{2}})} 
|z_{23}|^{2 ((h_{\tilde{l}_{1}}+n_{1}+1)-h_{l_{2}}-h_{l_{3}})}
|z_{13}|^{2 (h_{l_{2}}- (h_{\tilde{l}_{1}}+n_{1}+1)-h_{l_{3}})} \ .
\label{app:result_F}
\end{align}
To extract the corresponding minimal model
correlator~\eqref{app:odd3pt} we only have to shift the conformal
weights by the contribution $-\frac{m^{2}}{4 (k+2)}$ of the $U (1)_{2
(k+2)}$ part, 
\begin{align}
h_{l_{3}} &\to h_{l_{3}}-\frac{m_{3}^{2}}{4 (k+2)} = h_{l_{3},m_{3}}  \\
h_{l_{2}} &\to h_{l_{2}} - \frac{m_{2}^{2}}{4 (k+2)} = h_{l_{2},m_{2}} \\
h_{\tilde{l}_{1}} + n_{1}+1 &\to 
h_{\tilde{l}_{1}} + n_{1}+1 - \frac{(m_{1}-k-2)^{2}}{4 (k+2)} =
h_{l_{1},m_{1}} + \frac{1}{2} \ .
\end{align}
After simplifying the prefactor in~\eqref{app:result_F} we obtain our
final result for the minimal model correlator~\eqref{app:odd3pt},
\begin{align}
& \langle (\bar{G}^{+}_{-\frac{1}{2}}G^{+}_{-\frac{1}{2}}\phi_{l_{1},m_{1}}) (z_{1},\bar{z}_{1})
\phi_{l_{2},m_{2}} (z_{2},\bar{z}_{2})\phi_{l_{3},m_{3}}
(z_{3},\bar{z}_{3}) \rangle \nonumber\\
&\quad  = \frac{k+2}{2 (n_{1}+1) (l_{1}-n_{1})}  
\begin{pmatrix} \frac{l_{2}+m_{2}}{2} \\ n_{1}+1 \end{pmatrix}
\begin{pmatrix} \frac{l_{2}-m_{2}}{2}+n_{1}+1 \\ n_{1}+1\end{pmatrix}
\begin{pmatrix} l_{1} \\ n_{1}+1 \end{pmatrix}^{\!-1}\nonumber\\
&\qquad \times 
\begin{pmatrix}
\frac{\tilde{l}_{1}}{2} & \frac{l_{2}}{2} & \frac{l_{3}}{2} \\[1mm]
-\frac{\tilde{l}_{1}}{2} & \frac{m_{2}}{2}-n_{1}-1 & \frac{m_{3}}{2}
\end{pmatrix}^{\!2}
\sqrt{\big(\tilde{l}_{1}+1\big) \big(l_{2}+1\big) \big(l_{3}+1\big)}\, 
d_{\tilde{l}_{1},l_{2},l_{3}}\nonumber\\
&\qquad  \times 
|z_{12}|^{2
\left(h_{l_{3},m_{3}}-(h_{l_{1},m_{1}}+1/2)-h_{l_{2},m_{2}}\right)} 
|z_{23}|^{2 \left((h_{l_{1},m_{1}}+1/2)-h_{l_{2},m_{2}}-h_{l_{3},m_{3}}\right)}\nonumber\\
&\qquad \times 
|z_{13}|^{2 \left(h_{l_{2},m_{2}}-
(h_{l_{1},m_{1}}+1/2)-h_{l_{3},m_{3}}\right)}\ .
\label{app:oddcorrelatorresult}
\end{align}


\end{document}